\documentclass[preprint]{elsarticle}
\journal{European Journal of Mechanics/ B Fluids}
\usepackage{natbib}
\usepackage{geometry}
\usepackage{fleqn}
\usepackage{xcolor}
\usepackage{array}
\usepackage{bm}
\usepackage{graphicx}
\usepackage{caption}
\usepackage{amsmath}
\usepackage{physics}
\usepackage{subcaption}
\usepackage{float}
\usepackage{amsfonts}
\usepackage{enumitem}
\usepackage{multirow}

\biboptions{sort&compress}

\graphicspath{{./pictures/}}
\newcolumntype{\Lt}[1]{>{\raggedright\arraybackslash}p{#1}}
\newcolumntype{\Lc}[1]{>{\raggedright\arraybackslash}m{#1}}
\newcolumntype{\Cc}[1]{>{\centering\arraybackslash}m{#1}}

\begin{document}
\begin{frontmatter}
		
\title{Effect of Gaussian wake amplitude on wake-induced transition for a T106A low pressure turbine cascade}
		
		\author[1]{Aditi Sengupta\corref{corr-auth}}
		\ead{aditi@iitism.ac.in}
		
		\affiliation[1]{organization={Department of Mechanical Engineering, Indian Institute of Technology (Indian School of Mines)},%Department and Organization
			city={Dhanbad},
			postcode={826004}, 
			state={Jharkhand},
			country={India}}
		
		%%Research highlights
		%		\begin{highlights}
			%			\item 1
			%			\item 22
			%			\item 333
			%			\item 4444
			%			\item 55555
			%		\end{highlights}
		
\begin{abstract}
The wake-induced transition on the suction surface of a T106A low-pressure turbine (LPT) blade is investigated through a series of implicit large eddy simulations, solving the two-dimensional (2D) compressible Navier-Stokes equations (NSE). The impact of the incoming Gaussian wake amplitude on the blade's profile loss and associated boundary layer parameters is examined, revealing a 50\% reduction in skin friction drag at the highest amplitude. The results indicate that increasing wake amplitude leads to delayed separation and earlier reattachment, resulting in reduced separated flow. The vorticity and enstrophy dynamics during the transition process under varying wake amplitudes reveal characteristic features of wake-induced transition, such as puffs, streaks, and turbulent spots. The periodic passing of wakes induces intermittent \lq \lq calmed regions," which suppress flow separation and improve profile loss at low Reynolds numbers (Re), typically found in LPTs. The energy budget, accounting for both translational and rotational energy via the turbulent kinetic energy (TKE) and compressible enstrophy transport equation (CETE), respectively, shows trends with increasing wake amplitude. The relative contribution to TKE production and the roles of baroclinicity, compressibility, and viscous terms are explained.
\end{abstract}
		
%		\begin{keyword}
%			1 \sep 2 \sep 3 \sep 4 \sep 5 \sep 6
%		\end{keyword}
		
%		\newpageafter{abstract}	% Uncomment this in case you want abstract at new page
	\end{frontmatter}
	
	%% main text -------------------------------------------------------------
\section{Introduction}\label{sec1}

In modern high bypass ratio civil aircraft engines, the low-pressure turbine (LPT) contributes significantly to forward thrust, providing approximately 80\% of the power required for the fan and the first compressor stages \cite{opoka2007transition}. The efficiency of the LPT is directly linked to the specific fuel consumption efficiency and overall engine efficiency. For example, a 1\% improvement in LPT efficiency can result in a 0.5-1\% increase in fuel consumption efficiency, leading to an overall engine efficiency \cite{halstead1997boundary} improvement of about 0.7\%. Consequently, engine designers have concentrated on enhancing LPT efficiency, achieving levels \cite{sengupta2020numerical} of 90-93\%. However, further improvements are becoming increasingly challenging, prompting a shift towards \lq high-lift' blade designs. These designs feature large blade aspect ratios (3-7:1), resulting in primarily two-dimensional (2D) flow and subsonic Mach numbers in LPTs \cite{howell2002boundary}. This makes profile loss the largest contributor to total loss. The magnitude of profile loss is influenced by boundary layer development, with the suction surface accounting for 60\% of the total loss \cite{banieghbal1996wake}. Variations in boundary layer transition and separation processes can significantly affect profile loss for the same blade profile under different operating conditions \cite{lou2000separation}. Therefore, recent efforts to improve LPT loading and efficiency have focused on understanding and optimizing boundary layer transition and separation.

Boundary layers on LPT blades are usually transitional over a significant portion of the blade due to the low Reynolds numbers (Re) experienced by the low-pressure stages \cite{sengupta2020effects}. Additionally, these boundary layers are often unsteady, with LPTs experiencing incoming wakes from the upstream row of stator blades \cite{stieger2004transition}. The interaction of these wakes with downstream blade rows has been extensively studied \cite{wissink2006influence, lou2000separation, coull2012predicting, karaca2016dns}, highlighting an earlier transition to turbulence compared to natural transition due to the turbulence in the wakes. This increased turbulent flow over the blade surface leads to greater profile loss when boundary layers remain attached \cite{addison1990unsteady}, typically at higher Re and/or higher free stream turbulence intensities. However, at lower $Re$, where boundary layers tend to separate, the presence of upstream wakes and the resulting wake-induced transition can be beneficial \cite{sengupta2020effectsa}. Wakes have been found to reduce or suppress the growth of separation bubbles, thereby decreasing profile loss \cite{schulte1998unsteady, wissink2003dns}. This beneficial effect of suppressed separation regions on the suction surface has been observed not only in the presence of unsteady wakes but also with surface roughness \cite{sengupta2017roughness}, free stream excitation \cite{alam2000direct, wissink2004dns}, and even flutter phenomena \cite{sengupta2020effects, sengupta2020effectsa}. 

The bypass transition route, observed in the presence of free stream excitation \cite{sengupta2019direct, wissink2004dns} and roughness, has been extensively explored. The transition was shown to follow a nonmodal route \cite{sengupta2020nonmodal} with a sequence of unsteady separations along the suction surface of the blade. At low levels of free stream excitation, wake-induced transition involves an underlying Kelvin-Helmholtz mechanism \cite{watmuff1999evolution, sengupta2024separation}, where the turbulence of the wakes interacts, leading to the formation of \lq puffs', streaks, and turbulent spots \cite{wu1999simulation}. However, a parametric investigation of the role of wake amplitude (in particular) on the wake-induced transition mechanism has not been previously studied. Here, the separation suppression benefit afforded by wake-induced transition will be explored. We will determine whether wake amplitude is a suitable input for receptivity in the boundary layer and whether any improvement in profile loss can be obtained. Understandably, impinging wakes at the inflow are highly unsteady and the resultant flow dynamics are expected to be time-dependent too. However, most analyses focus on time-averaged boundary layer parameters to determine the role played by the wakes on the flow field. In addition to these traditional tools, here we will apply a compressible enstrophy equation (CETE) \cite{suman2022novel} directly on the instantaneous flow field. This approach will allow for an exploration of the role of wake amplitude on the individual budget terms that make up the CETE.

The discussion so far has been restricted to a model configuration \cite{wissink2004dns, sengupta2017roughness, alam2000direct}, which involves a streamwise pressure distribution imposed on a flat plate,  closely mimicking the suction surface of an actual LPT blade. The T106A blade profile, on the other hand, is a \lq real' \lq \lq first-generation" high-lift LPT blade, experiencing gradual loading compared to the model flat plate. The T106A profile has been studied experimentally \cite{stadtmuller2001test} and computationally \cite{wissink2003dns, sengupta2023compressibility, garai2015dns}. For a $Re$ of 51831, simulations of a T106A LPT passage \cite{ranjan2014direct} showed the presence of multiple unsteady separation bubbles along the suction surface and laminar-turbulent transition among many other intriguing fluid dynamical events. Direct numerical simulations (DNS) \cite{de2023effects} showed the role of stator and rotor interactions on the wake-induced transition evoked by periodic impact of unsteady wakes on the inflow to a T106A cascade. Another study on unsteady wakes in a T106A blade passage \cite{michelassi2015compressible}, explored the effect of turbulence levels in the background flow, the frequency of incoming wakes, and $Re$. The gap size between the stator and rotor of the T106A cascade was varied from 21.5\% to 43\% of the rotor chord length \cite{pichler2017highly}, revealing higher overall losses with a small gap. This was attributed to the amplification of incoming wake turbulent kinetic energy along the blade passage. In another study focusing on unsteady wakes impinging the T106A LPT profile \cite{lengani2019identification}, a proper orthogonal decomposition revealed that losses near the trailing edge of the suction surface were due to finer scales (higher order modes). These fine scales were integrated within the majority of the incoming wake. A recent study utilized spectral proper orthogonal decomposition to analyze flow characteristics in a T106A LPT passage cascade \cite{fiore2023t106} subject to upstream wakes. The analysis disclosed that specific modes were amplified within the suction side boundary layer, increasing turbulent structures in the developing wake downstream of the trailing edge. 

Prior investigations on T106A LPT blade profiles explored the impact of surface roughness, $Re$, free-stream turbulence, unsteady wakes with varying frequencies of wake passing, stator-rotors interactions, or various combinations thereof. To author's knowledge, there is limited research on the parametric influence of wake amplitude on the separation-induced transition process occurring on the suction surface of the T106A LPT blade, especially by exploring the enstrophy dynamics and the resultant energy budget. The current numerical investigation aims to address this by solving the two-dimensional (2D) compressible Navier-Stokes equations (NSE) for nine amplitudes of a periodic Gaussian wake imposed at the inflow. The aim is to examine how wake amplitude impacts the boundary layer thickness on the suction surface and, consequently, the profile loss \cite{denton1993loss}. To achieve this goal, we employ dispersion relation-preserving \cite{sagaut2023global} compact schemes as our numerical discretization method. These schemes are chosen for their ability to provide highly accurate simulations, ensuring the fidelity of our results. 

The structure of the paper unfolds as follows: The next section outlines the formulation of the problem slated for numerical simulation. This encompasses the governing equations and the boundary conditions prescribed for the nine simulated test cases. The adopted numerical methodology, along with previous validation endeavors \cite{sengupta2023compressibility}, is also described. Section \ref{sec3} encompasses the results and discussion, presenting instantaneous and time-averaged flow features in the presence of varying wake amplitude. The vorticity and enstrophy dynamics, its spectra, and the translational and rotational energy budgets are examined revealing insights into the wake-induced transition subject to different wake amplitudes. Bringing the paper to a conclusion, section \ref{sec4} provides a summary and the key conclusions drawn.

\section{Problem formulation of flow inside a T106A passage}\label{sec2}

\begin{figure}[ht!]
\centering
\includegraphics[width=\textwidth]{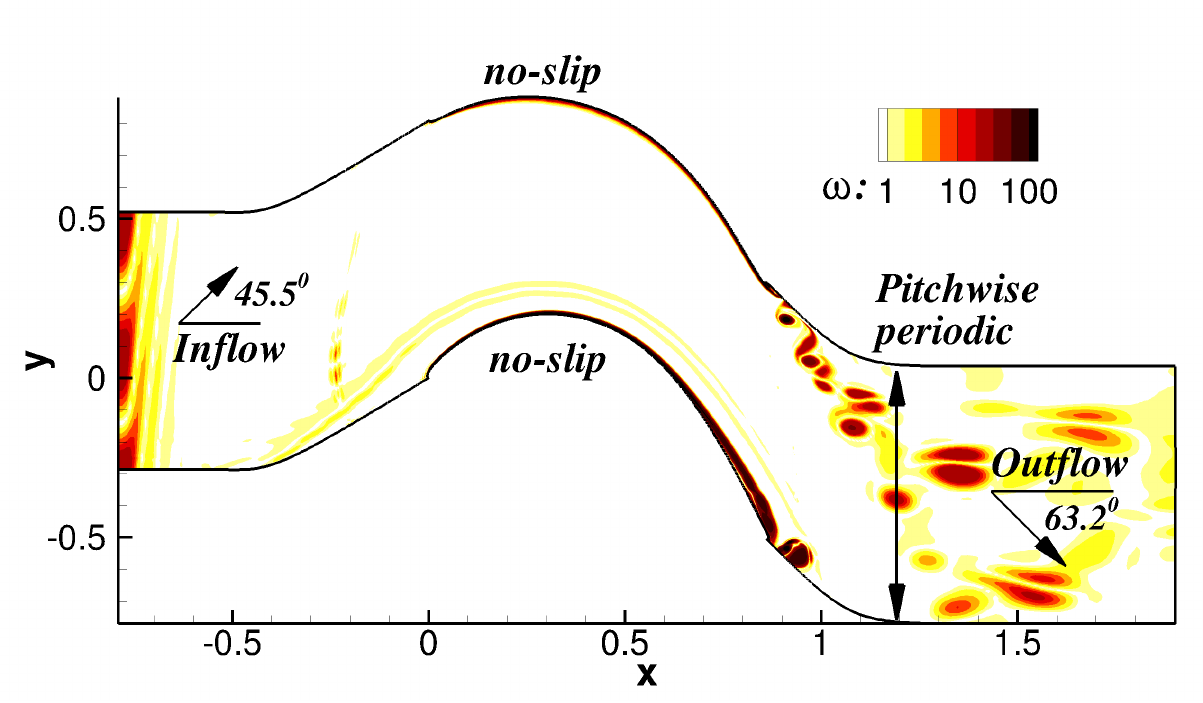}
\caption{Schematic of computational domain for wake-induced transition over a T106A LPT passage.}
\label{fig1}
\end{figure}

\noindent In a high-speed cascade wind tunnel at the Universität der Bundeswehr, München, Germany \cite{opoka2007transition, stadtmuller2001test}, a set of experiments were performed for a linear T106A cascade. Benchmark data sets for low $Re$ flows were obtained, which have been used for validating numerical simulations \cite{garai2015dns, sengupta2023compressibility}. The computational domain and boundary conditions of the present investigation, provided in Fig. \ref{fig1}, are based on this experimental configuration. The loading on the T106A blade is much more gradual than the contemporary \lq ultra high-lift' blade designs \cite{wissink2004dns, sengupta2017roughness}. The boundary layer developed on the suction surface of the T106A blade is much thinner, as a consequence.  A no-slip condition is imparted on the blade surfaces with periodicity ensured in the pitchwise direction, following the numerical setup of Wissink \cite{wissink2003dns}. Pitchwise periodicity is the commonly adopted boundary condition in previously reported works \cite{garai2015dns, wissink2006influence, michelassi2015compressible}. The inflow plane of the computational domain is placed upstream of the leading edge by a distance that is 0.4 times the axial chord length. The outflow is placed 0.5 times the axial chord length, downstream of the trailing edge. For meshing, a structured H-type grid is adopted which resolves the streamwise and wall-normal directions with 1251 and 501 points, respectively. The 1251 points in the streamwise direction are further subdivided into 276 points upstream of the leading edge, 751 points along the blade surface, and 226 points downstream of the blade. This has been done to ensure finer meshing on the surface of the LPT blades, to capture the small-scale turbulence induced by the imposed wakes. The simulations are conducted for a $Re$ of $1.6 \times 10^{5}$, based on the true chord $c$ and exit velocity $U_{TE}$. This follows the benchmark data sets in the previous DNS \cite{wissink2003dns} and experiments \cite{stadtmuller2001test}. The peak suction location for this mid-loaded T106A blade is at $42\% S_0$, where $S_0$ is the length of the suction surface. The pitch between blades is 0.9306. At the inflow (with an angle of attack, $\alpha_{in} = 45.5^0$), the inflow Mach number, $M_{s}$, is 0.15. The $Re$ at the inflow for the chosen length and time scales is calculated as 60142.72. The outflow has an angle of attack, $\alpha_{out} = -63.2^0$, and an outflow Mach number, $M_{out}$, of 0.404. To prevent spurious reflections from contaminating the inflow/outflow, a characteristic-based boundary condition using Riemann invariants \cite{hirsch1990numerical} is implemented.

We solve the unsteady 2D compressible NSE, ensuring the conservation of mass, momentum, and energy. These governing equations are provided in the divergence form \cite{sengupta2024thermal}, as it explicitly represents the conservation principles. This is crucial for accurately capturing physical phenomena, especially in problems involving shocks, discontinuities, or complex flow features \cite{hirsch1990numerical}. It also allows for the natural treatment of boundary conditions including incorporation of physical fluxes.

\begin{equation}
\frac{\partial \hat{Q}}{\partial t^*}+\frac{\partial \hat{E}}{\partial x^*}+\frac{\partial \hat{F}}{\partial y^*} =\frac{\partial \hat{E_v}}{\partial x^*}+\frac{\partial \hat{F_v}}{\partial y^*}
\label{ge1}
\end{equation}

\noindent where the conserved variables on the left hand side of Eq. \eqref{ge1} are given by

\begin{equation}
\hat{Q} = \left[ \rho^*; \; \rho^* u^*; \; \rho^* v^*; \; \rho^* e_t^* \right]^T
\label{ge2}
\end{equation}

\noindent The convective flux variables $\hat{E}$ and $\hat{F}$ are expressed as:

\begin{equation}
\hat{E} = \left[ \rho^* u^*; \; \rho^* u^{*2} + p^*; \;  \rho^* u^* v^*; \; (\rho^* e_t^* + p^*) u^*  \right]^T
\label{ge3}
\end{equation}

\begin{equation}
\hat{F} = \left[ \rho^* v^*; \; \rho^* u^* v^*; \; \rho^* v^{*2} + p^*; \; (\rho^* e_t^* + p^*) v^* \right]^T
\label{ge4}
\end{equation}

\noindent and the viscous flux vectors $\hat{E}_v$ and $\hat{F}_v$ are as follows:

\begin{equation}
\hat{E}_v = \left[ 0; \; \tau^*_{xx}; \; \tau^*_{xy}; \; u^* \tau^*_{xx} + v^* \tau^*_{xy} - q^*_x \right]^T
\label{ge6}
\end{equation}

\begin{equation}
\hat{F}_v = \left[ 0; \; \tau^*_{yx}; \; \tau^*_{yy}; \; u^* \tau^*_{yx} + v^* \tau^*_{yy} - q^*_y \right]^T
\label{ge7}
\end{equation}

\noindent The dimensional values of density, pressure, Cartesian components of fluid velocity, absolute temperature, and specific total energy of the fluid are symbolized as $\rho^*$, $p^*$, $u^*$, $v^*$, $T^*$, and $e_t^*$, respectively. As the working fluid is air for the present numerical investigation, the specific heat ratio, denoted by $\gamma$, is prescribed a value of 1.4. The components of symmetric viscous stress tensor, expressed by $\tau^*_{xx}$, $\tau^*_{yy}$, $\tau^*_{xy}$, $\tau^*_{yx}$, are related to the velocity gradients as, 

\begin{equation}
\tau^*_{xx} = \biggl( 2 \mu \frac{\partial u^*}{\partial x^*} + \lambda \nabla^* \cdot \vec{V}^* \biggr ), \; \tau^*_{yy} = \biggl( 2 \mu \frac{\partial v^*}{\partial y^*} + \lambda \nabla^* \cdot \vec{V}^* \biggr ), 
\label{ge9}
\end{equation}

\begin{equation}
\tau^*_{xy} = \tau^*_{yx} = \mu \biggl( \frac{\partial u^*}{\partial y^*} +  \frac{\partial v^*}{\partial x^*} \biggr )
\label{ge10}
\end{equation}

\noindent Fluid properties of specific heat capacity ($C_v$) and thermal conductivity ($\kappa$) are treated as constants in this formulation. The heat conduction terms $q^*_x$ and $q^*_y$ are expressed as,

\begin{equation}
q^*_x = - \kappa \frac{\partial T^*}{\partial x^*}, \; q^*_y = - \kappa \frac{\partial T^*}{\partial y^*}
\label{ge11}
\end{equation}

\noindent The coefficient of molecular viscosity, $\mu$, is related to the second coefficient of viscosity, $\lambda$, through Stokes' hypothesis, which is expressed as $\lambda = -2\mu/3$. This is a commonly employed constitutive relation. Sutherland's law is applied to model the viscosity as a function of temperature. The system of equations is closed using the ideal gas law which is expressed as follows,

\begin{equation}
p^*=\rho^* R^*T^*
\label{ge12}
\end{equation}

\noindent which in turn defines the specific energy, $e_t^*$ as follows,

\begin{equation}
e_t^* =\frac{p^*}{[\rho^*(\gamma-1)]} + \frac{(u^{*2} + v^{*2})}{2}
\label{ge13}
\end{equation}

\noindent Non-dimensionalization is performed using the scales of temperature and density, chosen here as follows: $T_s = 300K$ and $\rho_s = 1.177 kg/m^3$. To investigate the wake-induced transition on the suction surface of the T106A blade passage, nine numerical simulations are conducted with varying amplitudes of the Gaussian wake imposed at the inflow plane. Each simulation requires computations with 80 cores over 175 compute hours, totaling to 14,000 core hours per test case. Initial transients in the flow are circumvented by simulating five to six through-flows. For time-averaging the flow field, an additional 10-12 through-flows are computed, and the time-resolved data has been stored at an interval of non-dimensional time, $t = 0.05$.

\subsection{Implementing unsteady Gaussian wakes at the inflow}
The effect of unsteady wakes was introduced through a sinusoidal perturbation to the free-stream flow in an earlier experimental study \cite{lou2000separation}. In our current investigation of wake-induced transition, we will use a similar approach, employing a periodic Gaussian function to introduce unsteadiness into the free-stream flow. This methodology is consistent with the procedures used in previous studies \cite{karaca2016dns, sengupta2020effectsa}. We introduce a Gaussian wake at the inflow using the following velocity profile, which is added to the streamwise inlet velocity of Eq. \eqref{ge1} during each time step. The expression for the Gaussian wake's velocity profile is given as,

\begin{equation}
u(w) = a_{wake} e^{-\alpha  \biggl [ mod \biggl ( \frac{y}{y_{max}} + \frac{2t}{t_{wake}} - 1, 2 \biggr ) \biggr ]^2}
\label{wake}
\end{equation}

\noindent Here, $a_{wake}$ is the amplitude of the Gaussian wake, $\alpha$ is the exponential constant prescribing width of the Gaussian profile, $y_{max}$ is the maximum wall-normal height of the computational domain shown in Fig. \ref{fig1}, and $t_{wake}$ denotes the time period of wake passing. Equation \eqref{wake} ensures that during each wake passing cycle, the incoming wake takes into account the relative position of the LPT blade with respect to upstream stator blades by including the associated phase shift \cite{sengupta2020numerical}. For the present numerical simulations, we set the non-dimensional time period of wake passing to $t_{wake} = 0.35$, and the exponential constant of the Gaussian function to $\alpha = 19$, aligning with wake spacing considered by Karaca and Gungor \cite{karaca2016dns}. We must also ensure that the parameters allow one to remain within acceptable limits of the non-dimensional Strouhal number ($St \approx 0.3$) for upstream blade-induced wakes \cite{sengupta2020numerical}. The $St$ is defined as $St = L_s / (U_s t_{wake})$, where $L_s$ and $U_s$ represent the length and velocity scales, respectively. These are chosen as the chord length and the velocity at the trailing edge, rendering $St = 0.3403$. To assess the influence of wake amplitude, nine simulations were conducted with $a_{wake}$ ranging from 0.1 to 0.9.

\begin{figure}[ht!]
\centering
\includegraphics[width=.9\textwidth]{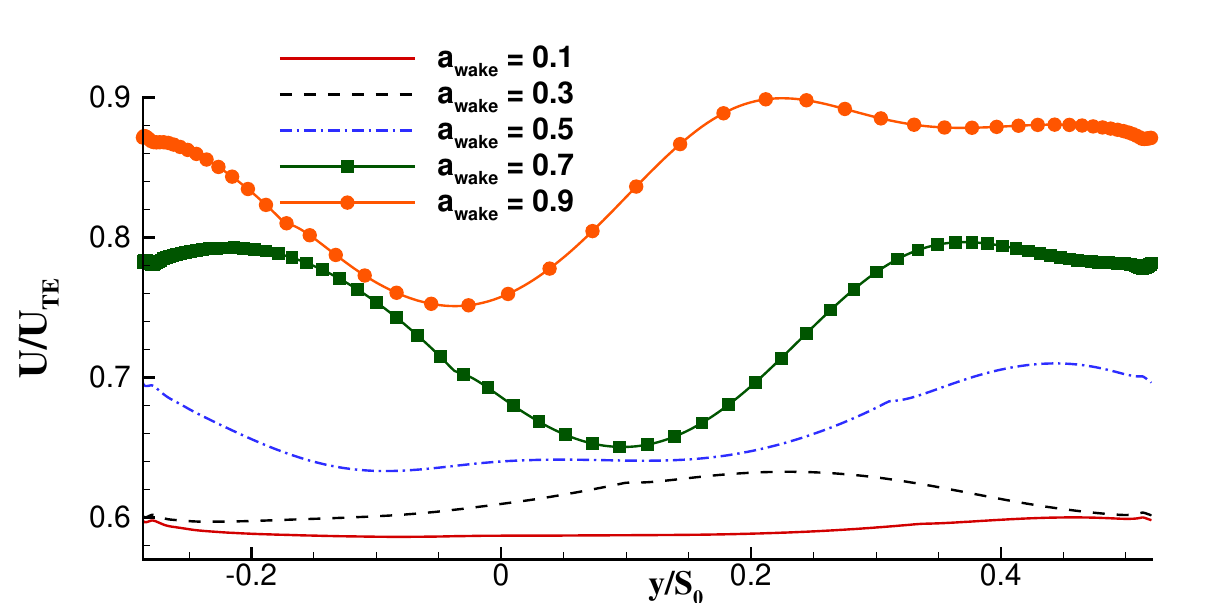}
\caption{Time-averaged mean velocity at the inflow plane of the T106A LPT passage for imposed wake amplitude, $a_{wake}$ = 0.1, 0.3, 0.5, 0.7 and 0.9.}
\label{fig2}
\end{figure}

The inflow characteristics of the imposed Gaussian wake are examined for the different values of $a_{wake}$ by comparing the normalized time-averaged streamwise velocity, extracted at the inflow plane in Fig. \ref{fig2}. The deficit in the velocity profile due to the wake is evident, particularly for $a_{wake} = 0.7$ and 0.9. For $a_{wake} = 0.9$, the deficit is approximately 10\% of the mean velocity, whereas for $a_{wake} = 0.7$, it is slightly lower (9\% of the mean velocity). The location where this deficit occurs also shifts upstream with the higher amplitude. Wakes are known to impart a negative jet in the flow from the free stream. This transports turbulent fluid from the wake to the edge of the boundary layer \cite{addison1990unsteady}, and it can be seen from Fig. \ref{fig2} that this tendency increases for higher $a_{wake}$. For $a_{wake} < 0.5$, the velocity deficit (ranging from 2\% to 5\% of the mean velocity) is much lower in amplitude and it is expected that it will have a reduced impact on the separation-induced transition on the suction surface.

\subsection{Numerical methodology} 
\noindent In our simulations, we utilize a dispersion relation preserving (DRP) compact scheme for spatial discretization, optimized for neutral stability, as outlined by Sagaut {\it et al.} \cite{sagaut2023global}. To calculate the convective derivative, we employ a second-order optimized upwind compact scheme (OUCS3), which ensures that numerical and physical dispersion relations match across a wide range of wavenumbers and frequencies. This match is particularly important for the wave-dominated flows observed on the suction and pressure surfaces of the LPT blade \cite{sengupta2020effects}. The numerics significantly influence the series of unsteady separation bubbles obtained for pressure gradient dominated internal flows \cite{sengupta2020nonmodal}. To ensure numerical stability, a one-dimensional fifth-order filter with a coefficient of 0.47 is applied in both $x$ and $y$-directions. For calculating viscous derivatives, we use the second-order central difference scheme. Time integration is carried out using the OCRK3 scheme, an optimized version of the three-stage Runge-Kutta method, which offers better DRP properties compared to the classical four-stage, fourth-order Runge-Kutta method \cite{sagaut2023global}. The chosen time step for the simulations is $5 \times 10^{-6}$, selected to maintain neutral stability of the numerical methods used for discretization.

The numerical framework used in this study has been benchmarked against an experimental study \cite{stadtmuller2001test} and the DNS of Wissink {\it et al.} \cite{wissink2003dns} for the T106A cascade, as detailed in previously reported work \cite{sengupta2023compressibility}. The baseline flow, devoid of any free-stream disturbances, was accurately modeled, with the time-averaged distribution of the pressure coefficient, $C_p$ showing a good match between the computed and benchmark values. This validation confirmed that the essential flow phenomena, including separation on both the suction and pressure surfaces, were accurately captured.

\section{Results and Discussion}\label{sec3}

In this section, we will discuss the time-averaged flow field for varying inflow wake amplitude, particularly emphasizing the boundary layer parameters and separation events on the suction surface of the T106A LPT blade, as these have a potential impact on the profile loss experienced by the blade. Furthermore, we will explore the instantaneous rotationality of the flow field as a function of the inflow wake amplitude by exploring the vorticity and enstrophy dynamics. Lastly, we will present the energy budget via the turbulent kinetic energy and compressible enstrophy, so that insights can be drawn about production of turbulence and the various mechanisms responsible for the wake-induced transition in an LPT passage.

\subsection{Role of wake amplitude on time-averaged flow}

\begin{figure}[ht!]
\centering
\includegraphics[width=.9\textwidth]{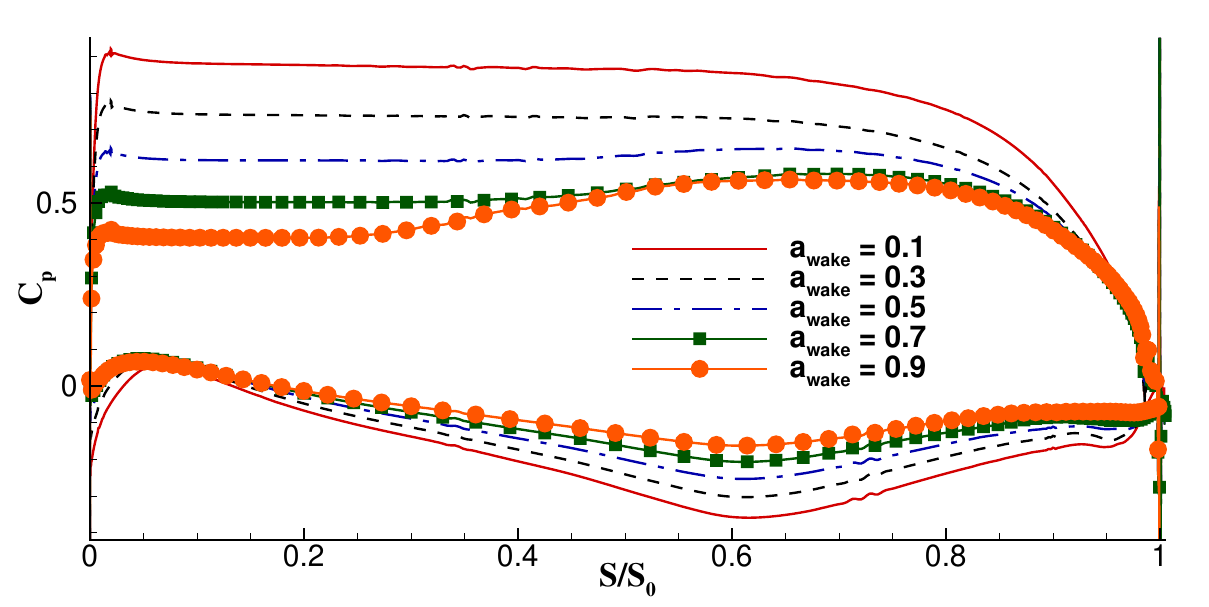}
\caption{Streamwise variation of time-averaged $C_p$, for inflow wake amplitudes, $a_{wake}$ = 0.1, 0.3, 0.5, 0.7 and 0.9.}
\label{fig3}
\end{figure}

\noindent The flow is allowed to develop for five through flows before we begin gathering the statistics for the time-averaging. This has been done to ensure that any early-time transients are flushed off. The time-averaged streamwise variations of the coefficient of static pressure, $C_p$ are shown for the indicated wake amplitudes in Fig. \ref{fig3}. In the neighborhood of the trailing edge on the suction side, a jump is noted in the gradient of the pressure coefficient, which is sensitive to the amplitude of the incoming wakes. For $a_{wake} = 0.1$, this jump occurs at $S/S_0 \approx 0.93$, which is consistent with prior research \cite{wissink2003dns} exploring role of periodic Gaussian wakes at the inflow of an LPT cascade. As the wake amplitude is increased, the location along the suction surface for this jump in $C_p$ shifts downstream. For example, for $a_{wake} = 0.3$, the jump occurs at $S/S_0 \approx 0.9$. Interestingly, the separation bubble near the leading edge is suppressed with increasing $a_{wake}$, and beyond $a_{wake} = 0.5$, this separation bubble no longer exists. The long flat plateau near the trailing edge on the suction surface is also indicative of a separation bubble \cite{ranjan2014direct}. The downstream flow on the suction surface is found to be receptive to the pressure distribution at leading edge. With increasing $a_{wake}$, the extent of the separation bubble at the trailing edge is reduced. For $a_{wake} >$ 0.5, a noticeable lack of the sharp jump in the pressure distribution near the trailing edge, suggests a suppression of the separated flow with increasing wake amplitude.  

On the pressure surface in Fig. \ref{fig3}, a stagnation point is observed which causes the flow to be redirected around the blade's nose, to overcome the adverse pressure gradient \cite{michelassi2015compressible}. For cases with lower $a_{wake}$, the pressure is relatively low in the vicinity of the stagnation point, which combined with the curvature of the LPT blade can cause flow reversal, particularly when the magnitude of the adverse pressure gradient is high. This is found to be the case for uniform inflow \cite{garai2015dns}, but in the presence of incoming wakes, the adverse pressure gradient is much weaker, particularly at higher wake amplitudes, thus the tendency for the flow to separate on the pressure surface decreases further. 

\begin{figure}[ht!]
\centering
\includegraphics[width=.9\textwidth]{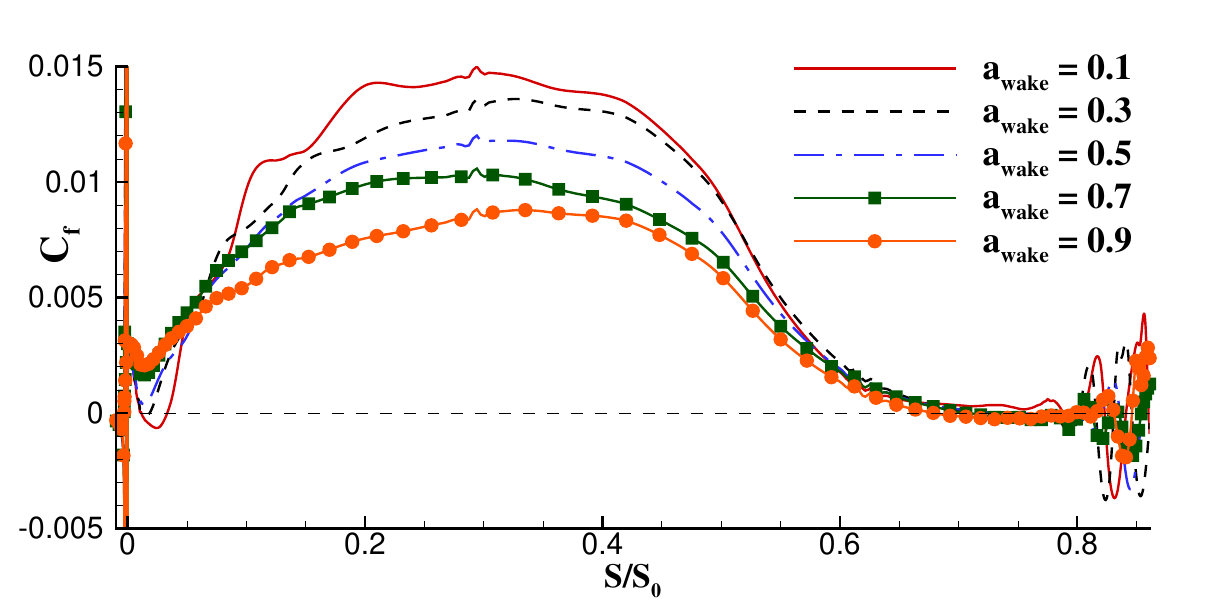}
\caption{Streamwise variation of time-averaged skin friction coefficient, $C_f$, for inflow wake amplitudes, $a_{wake}$ = 0.1, 0.3, 0.5, 0.7 and 0.9 for wake-induced transition over the suction surface of a T106A LPT blade.}
\label{fig4}
\end{figure}

In Fig. \ref{fig4}, the streamwise distribution of the time-averaged skin friction coefficient, $C_f$ is shown for the indicated amplitudes of the incoming wakes. For steady separations, the zero crossings of the $C_f$ plot are considered as approximate locations of flow separation ($S_{sep}$) and reattachment ($S_{reatt}$), while the location along the suction surface where the global minimum exists is identified as the transition location ($S_{trans}$) \cite{sengupta2020numerical}. Two separation bubbles are noted: one near the leading edge (at $S/S_0 \approx 0.05$) and the other near the trailing edge. With an increase in wake amplitude, delayed separation is observed with an earlier reattachment and transition, thereby bringing down the effective separated region. For $a_{wake} > 0.5$, the leading edge bubble is completely suppressed whereas the streamwise extent of the separation bubble at the trailing edge is significantly truncated compared to the scenario for $a_{wake} = 0.1$. The skin friction decreases with increasing $a_{wake}$, dropping by approximately 50\% for $a_{wake} = 0.9$ compared to $a_{wake} = 0.1$. Thus, introducing wakes at the inflow potentially reduces skin friction drag along the suction surface boundary layer. The effect of increasing $a_{wake}$ parallels the impact of increasing Mach number \cite{sengupta2023compressibility} or turbulent intensity of free stream turbulence \cite{sengupta2024separation}, but with a more pronounced influence on reducing the surface area exposed to flow separation and skin friction drag.

Table \ref{tab1} documents locations of separation, flow transition, and reattachment for the two observed separation bubbles shown in Fig. \ref{fig4} across nine different amplitudes of incoming wakes. As discussed, increasing wake amplitude suppresses separation events, leading to delayed separation and earlier reattachment. The accelerated transition of the flow with increasing wake amplitude can be attributed to the \lq negative jet' effect, where turbulent fluid from the wake influences the boundary layer edge \cite{addison1990unsteady}. For $a_{wake} > 0.4$, the leading edge bubble is completely suppressed. Meanwhile, for the separation bubble near the trailing edge, its streamwise extent decreases by 23.3\% for $a_{wake} = 0.9$, compared to $a_{wake} = 0.1$. 

\begin{table}[h!]
\centering
\caption{Locations of separation, transition, and reattachment at the leading and trailing edges for different inflow wake amplitudes.}
\vspace{1mm}
\begin{tabular}{|c| c| c| c| c| c| c| c|}
\hline
 Case & LE: $S_{sep}$ & LE: $S_{trans}$ & LE: $S_{reatt}$ & TE: $S_{sep}$ & TE: $S_{trans}$ & TE: $S_{reatt}$ \\ [0.5ex]
 \hline\hline
    $a_{wake}$ = 0.1 & 0.0106 & 0.0238 & 0.0333 & 0.7889 & 0.8309 & 0.8539\\ 
    \hline
    $a_{wake}$ = 0.2 & 0.0108 & 0.0158 & 0.0217 & 0.7771 & 0.8301 & 0.8372 \\ 
    \hline
    $a_{wake}$ = 0.3 & 0.0121 & 0.0152 & 0.0187 & 0.7802 & 0.8233 & 0.8310 \\
    \hline
    $a_{wake}$ = 0.4 & - & - & - & 0.7814 & 0.8246 & 0.8312 \\
    \hline
    $a_{wake}$ = 0.5 & - & - & - & 0.8001 & 0.8120 & 0.8212 \\
    \hline
    $a_{wake}$ = 0.6 & - & - & - & 0.8245 & 0.8306 & 0.8408 \\
    \hline
    $a_{wake}$ = 0.7 & - & - & - & 0.8292 & 0.8372 & 0.8461 \\
    \hline
    $a_{wake}$ = 0.8 & - & - & - & 0.8281 & 0.8375 & 0.8436 \\
    \hline
    $a_{wake}$ = 0.9 & - & - & - & 0.8308 & 0.8395 & 0.8460 \\
    [1ex]
 \hline
\end{tabular}

\label{tab1}
\end{table}

\begin{figure}[ht!]
\centering
\includegraphics[width=.9\textwidth]{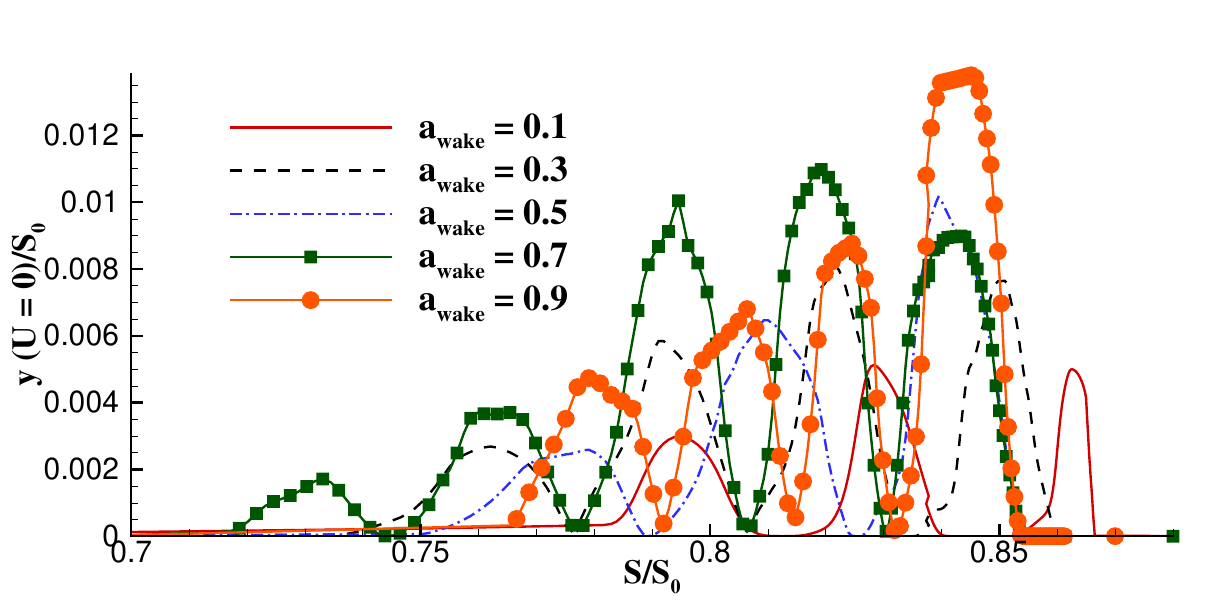}
\caption{Time-averaged separation bubble half heights for inflow wake amplitudes, $a_{wake}$ = 0.1, 0.3, 0.5, 0.7 and 0.9, on suction surface of T106A LPT blade.}
\label{fig5}
\end{figure}

Figure \ref{fig5} illustrates the time-averaged half-heights of unsteady separation bubbles along the suction surface of the T106A blade under varying amplitudes of incoming Gaussian wakes. For $a_{wake} = 0.1$, three unsteady separation bubbles are observed along the suction surface, consistent with previous simulations \cite{ranjan2014direct, garai2015dns, michelassi2015compressible} of the T106A passage with uniform inflow. As wake amplitude increases, the number of separation bubbles also increases. For example, at $a_{wake} = 0.3$ and 0.5, four separation bubbles are observed. Furthermore, increasing $a_{wake}$ leads to larger bubble half-heights, but their streamwise extent decreases, resulting in more localized bubbles. At  $a_{wake} = 0.9$, the maximum wall-normal extent reaches $y/S_0 = 0.0135$, whereas for $a_{wake} = 0.1$, the maximum height is 0.005. This represents a 37\% increase in bubble height. In the instantaneous flow field, these unsteady separation bubbles merge and spawn from within the suction surface boundary layer, highlighting the integrated effect of higher wake amplitudes on net separated flow. This, in turn, influences boundary layer characteristics and profile losses \cite{banieghbal1996wake}, as discussed next. The role of wake amplitude in influencing the number and extent of separation bubbles in both streamwise and wall-normal directions mirrors the impact of Mach number \cite{sengupta2023compressibility} and turbulent intensity of free-stream turbulence \cite{sengupta2024separation}. However, wake amplitude exhibits a more pronounced effect on the suction surface boundary layer compared to Mach number or turbulent intensity.

\begin{figure}[ht!]
\centering
\includegraphics[width=.75\textwidth]{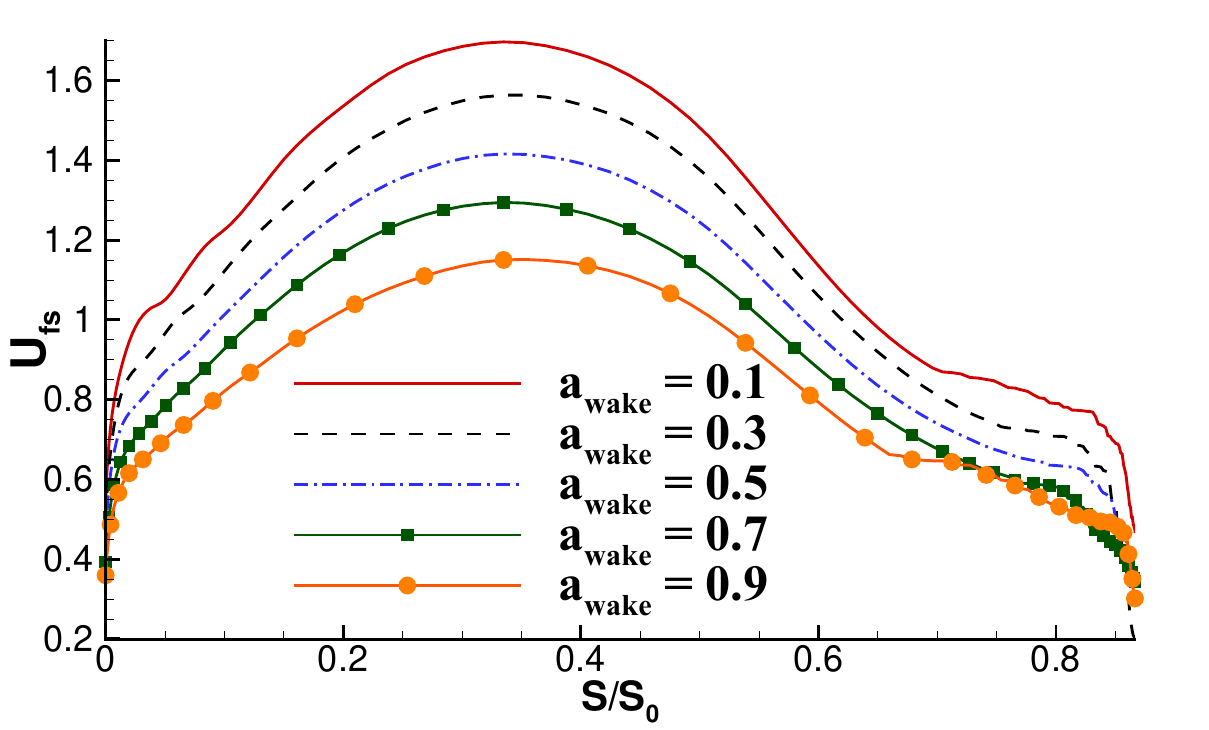}
\caption{Time-averaged free stream velocity at edge of the boundary layer on the suction surface, $U_{fs}$ for inflow wake amplitudes, $a_{wake}$ = 0.1, 0.3, 0.5, 0.7 and 0.9.}
\label{fig6}
\end{figure}

Figure \ref{fig6} presents the time-averaged free stream velocity at the edge of the boundary layer, $U_{fs}$, for various inflow wake amplitudes. Across all wake amplitudes, the velocity distribution along the suction surface reveals an accelerated pre-transitional boundary layer where a favorable pressure gradient exists (for $S/S_0 < 0.4$). This acceleration continues until reaching a peak velocity, typically occurring slightly upstream of the peak suction location at $S/S_0 = 0.42$. The magnitude of this peak velocity decreases with increasing $a_{wake}$, similar to trends observed with increasing Mach number \cite{sengupta2023compressibility} or turbulent intensity \cite{sengupta2024separation}, but with a more pronounced deviation from cases with uniform inflow. Beyond $S/S_0 = 0.42$, the flow decelerates as the adverse pressure gradient becomes insurmountable, leading to boundary layer separation. Once the separated shear layer forms, the flow remains stagnant beneath it, resulting in a plateau of constant velocity (for $0.65 < S/S_0 < 0.8$). With increasing wake amplitude, the extent of this velocity plateau decreases \cite{coull2012predicting}, indicating suppression of separated flow. Further downstream, a rapid deceleration is observed, followed by flow reattachment ($S/S_0 \approx 0.8$), occurring more rapidly at lower values of $a_{wake}$.

The blade's profile loss is often approximated in terms of the total pressure loss downstream of the blade row in an LPT cascade \cite{denton1993loss}, which is expressed as the following:

\begin{equation}
\zeta_p = \biggl ( \frac{-C_{pb} t_{TE}}{p_b cos\alpha_{ex}} \biggr ) + \biggl ( \frac{ \delta^{TE} + t_{TE}}{p_b cos\alpha_{ex}} \biggr )^2 + \biggl ( \frac{2 \theta_{TE}}{p_b cos\alpha_{ex}} \biggr )
\label{mom_loss}
\end{equation}

\begin{figure}[ht!]
\centering
\includegraphics[width=.75\textwidth]{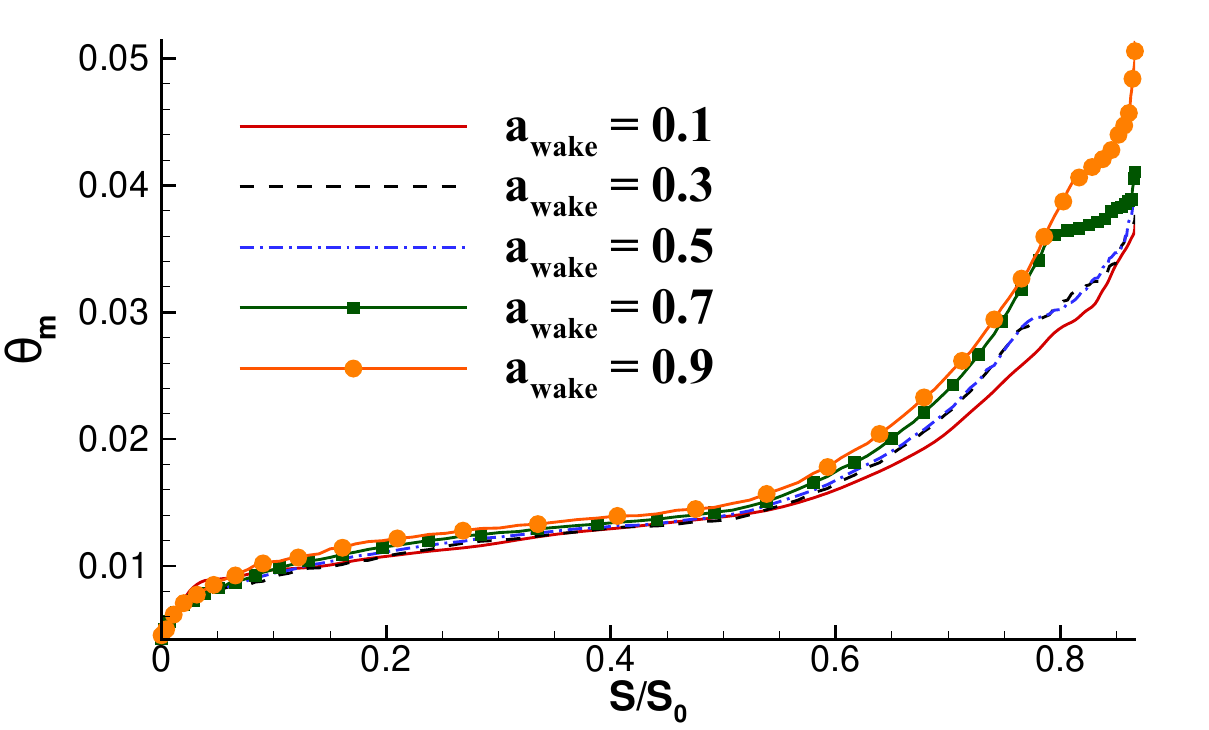}
\caption{Time-averaged momentum thickness, $\theta_{m}$ for inflow wake amplitudes, $a_{wake}$ = 0.1, 0.3, 0.5, 0.7 and 0.9, on the suction surface of a T106A LPT blade.}
\label{fig7}
\end{figure}

Here, the base pressure coefficient is symbolized by $C_{pb}$, the thickness at the trailing edge is given by $t_{TE}$, $p_b$ is the pitch between the blades and $\alpha_{ex}$ is the exit flow angle relative to axial direction. The parameters, $\delta_{TE}$ and $\theta_{TE}$ represent the displacement and momentum thickness at the trailing edge, respectively. The contributions to the total loss of the three terms on the right hand side of Eq. \eqref{mom_loss} are as follows: (i) first term which accounts for loss due to low base pressure acting at trailing edge has a meager 3\% contribution, (ii) second term which is proportional to the square of the displacement thickness, accounts for a 7\% contribution. This term arises due to blockage offered by both boundary layer development and trailing edge thickness. (iii) The third and largest contributor to the total loss (a swooping 90\%) is due to mixed-out loss of boundary layers. Here too, the term is directly proportional to the trailing edge boundary layer thickness. Thus, we depict the streamwise distribution of the time-averaged momentum thickness along the suction surface as a function of the wake amplitude in Fig. \ref{fig7}, to comment on its contribution to the blade profile loss. In this calculation, we use the minimum vorticity threshold for identifying the edge of the boundary layer.

Irrespective of the magnitude of $a_{wake}$, the maximum momentum thickness is obtained at the trailing edge. Upon increasing the wake amplitude from 0.1 to 0.9, the momentum thickness is increased by 39.2\%. When the Mach number was doubled \cite{sengupta2023compressibility}, the momentum thickness was found to increase by 67.7\% whereas when the turbulent intensity was increased from 0\% to 7\%, the momentum thickness was found to increase \cite{sengupta2024separation} only by 6.25\%. This highlights that increasing compressibility has a much more pronounced effect on profile loss compared to increasing wake amplitude. On the other hand, increasing free stream turbulence intensity has a marginal effect on profile loss. While the benefit of separation suppression of wake-induced transition is less than that of free stream turbulence, we also observed a significant reduction in skin friction drag by 50\% for the highest $a_{wake}$, whereas skin friction drag increased with turbulent intensity of free stream turbulence \cite{sengupta2024separation}. Overall, in terms of both skin friction drag and pressure drag experienced on the suction surface of the LPT blade, high-amplitude Gaussian wakes at the inflow provide a beneficial effect compared to both increased compressibility and free stream turbulence. 

Table \ref{tab2} summarizes the impact of increasing wake amplitude on time-averaged boundary layer parameters on the suction surface, including maximum velocity at the boundary layer edge, momentum thickness, and bubble half-height. Increasing wake amplitude reduces the adverse pressure gradient, resulting in a decrease in peak velocity magnitude and its occurrence further downstream along the suction surface. As discussed in relation to Fig. \ref{fig5}, higher $a_{wake}$ leads to increased number and heights of unsteady separation events, although these bubbles are localized, reducing overall separated flow. For instance, the height of the separation bubble increases from $y_{bub} = 0.0051$ to 0.0138, when wake amplitude is increased ninefold. Similarly, the momentum thickness at the trailing edge increases by 1.4 times when increasing wake amplitude from 0.1 to 0.9.

\begin{table}[h!]
\centering
\caption{Maximum free stream velocity at edge of boundary layer ($U_{fs}$), momentum thickness ($\theta_{m}$), and bubble half height ($y_{bub}$) for different inflow wake amplitudes.}
\vspace{1mm}
\begin{tabular}{|c| c| c| c|}
\hline
 Case & Max $(U_{fs})$ & Max $(\theta_{m})$ & Max ($y_{bub}$) \\ [0.5ex]
 \hline\hline
    $a_{wake}$ = 0.1 & 1.6971 & 0.0369 & 0.0051 \\ 
    \hline
    $a_{wake}$ = 0.2 & 1.6304 & 0.0372 & 0.0066 \\ 
    \hline
    $a_{wake}$ = 0.3 & 1.5637 & 0.0377 & 0.0081 \\
    \hline
    $a_{wake}$ = 0.4 & 1.4899 & 0.0384 & 0.0094 \\
    \hline
    $a_{wake}$ = 0.5 & 1.4161 & 0.0391 & 0.0101 \\
    \hline
    $a_{wake}$ = 0.6 & 1.3553 & 0.0401 & 0.0104 \\
    \hline
    $a_{wake}$ = 0.7 & 1.2946 & 0.0411 & 0.0109 \\
    \hline
    $a_{wake}$ = 0.8 & 1.2231 & 0.0462 & 0.0124 \\
    \hline
    $a_{wake}$ = 0.9 & 1.1515 & 0.0514 & 0.0138 \\
    [1ex]
 \hline
\end{tabular}

\label{tab2}
\end{table}

\subsection{Creation and distribution of rotationality during wake-induced transition}

Rotationality affects the aerodynamic behavior of the flow through the LPT blade passages. Vorticity is a measure of the local rotation of fluid elements \cite{doering1995applied} and the creation of vortices leads to more complex pressure and velocity distributions within the fluid. It is closely related to the transfer of energy within a fluid via the quantity, enstrophy. Enstrophy, which is defined as the integral of the square of the vorticity over a given volume, provides a measure of the intensity of rotational motion within a fluid. High enstrophy indicates strong rotational motions and a high concentration of vortices, which are essential features in many fluid flows, particularly turbulent ones. Regions with high enstrophy can indicate areas of potential instability and transition to turbulence \cite{sengupta2018enstrophy}. In turbulent flows, rotationality is a key mechanism by which energy cascades from larger to smaller scales. A proper understanding of the creation and propagation of rotationality in the flow can aid in designing blade profiles that maximize energy extraction and minimize losses due to flow separation and turbulence. We intend to examine the vorticity and enstrophy of the instantaneous flow field in this section to extract the importance of wake amplitude on the vorticity generation within the LPT passage.

\begin{figure}[ht!]
\centering
\includegraphics[width=\textwidth]{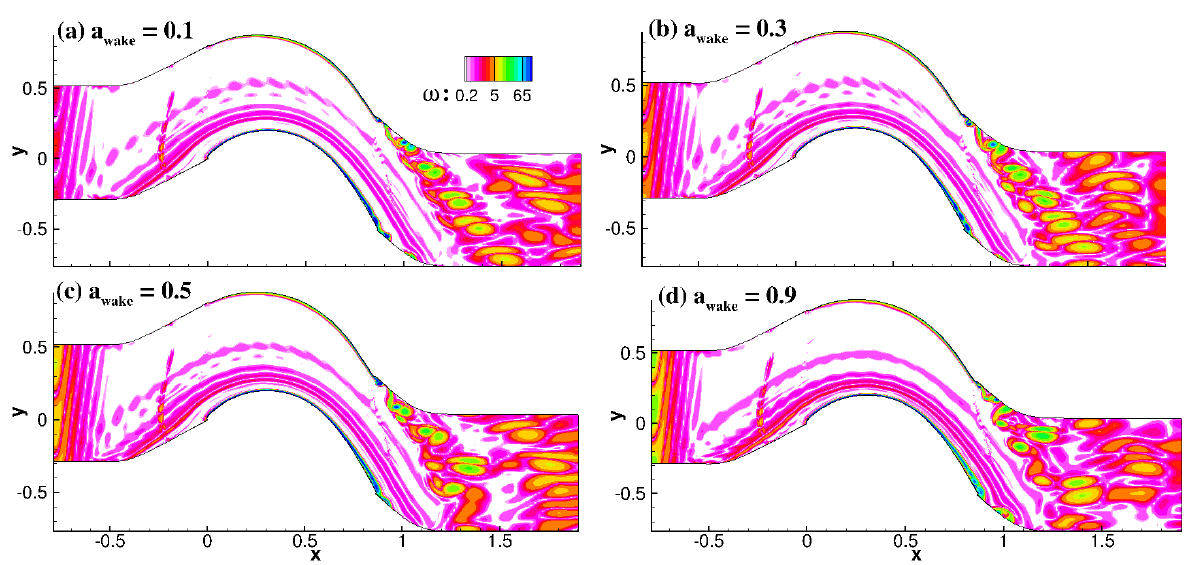}
\caption{Contours of $|\omega|$ for inflow wake amplitudes, (a) $a_{wake}$ = 0.1, (b) $a_{wake}$ = 0.3, (c) $a_{wake}$ = 0.5 and (d) $a_{wake}$ = 0.9.}
\label{fig8}
\end{figure}

In Fig. \ref{fig8}, the contours of vorticity magnitude, $|\omega|$ are compared for the indicated inflow wake amplitudes. On the suction surface, a thin boundary layer develops starting from the leading edge as this is the key receptivity site due to low $Re$. Beyond the peak suction location ($S/S_0 = 0.42$), the flow separates due to the adverse pressure gradient. The breakdown to turbulence results in reattachment of the separated shear layer enclosing unsteady separation bubbles. These are seen on the aft portion of the blade, near the trailing edge. The \lq puff'-like structures spawning from the inflow plane are characteristic of wake-induced transition \cite{wissink2006influence}. In the free stream, a stream of blobs of vorticity are noted which are representative of the small-scale turbulence in the wake \cite{stieger2004transition}. With an increase in $a_{wake}$, the thickness of the boundary layer and the number of separation bubbles increase. The vorticity in the close vicinity of the suction surface is found to decrease in magnitude with an increase in $a_{wake}$, suggesting a redistribution of vorticity over a larger number of separation bubbles. The incoming wakes at the inflow plane also produce more vorticity for higher $a_{wake}$. The blobs of vorticity in the free stream coalesce into each other with an increase in $a_{wake}$, eventually merging into a large-scale line vortex for $a_{wake} = 0.9$ in Fig. \ref{fig8}(d). 

On the pressure surface, there is no flow separation, instead vortex shedding occurs from the trailing edge. With an increase in $a_{wake}$, the intensity of vortex shedding is reduced, due to the influence of the wake on the pressure distribution (as noted in Fig. \ref{fig3}). The flow appears to be more chaotic due to vortices in the free stream (created by the imposed wakes) and the isolated wake components penetrating inside the boundary layer due to a shear sheltering effect \cite{jacobs2001simulations}.

\begin{figure}[ht!]
\centering
\includegraphics[width=.85\textwidth]{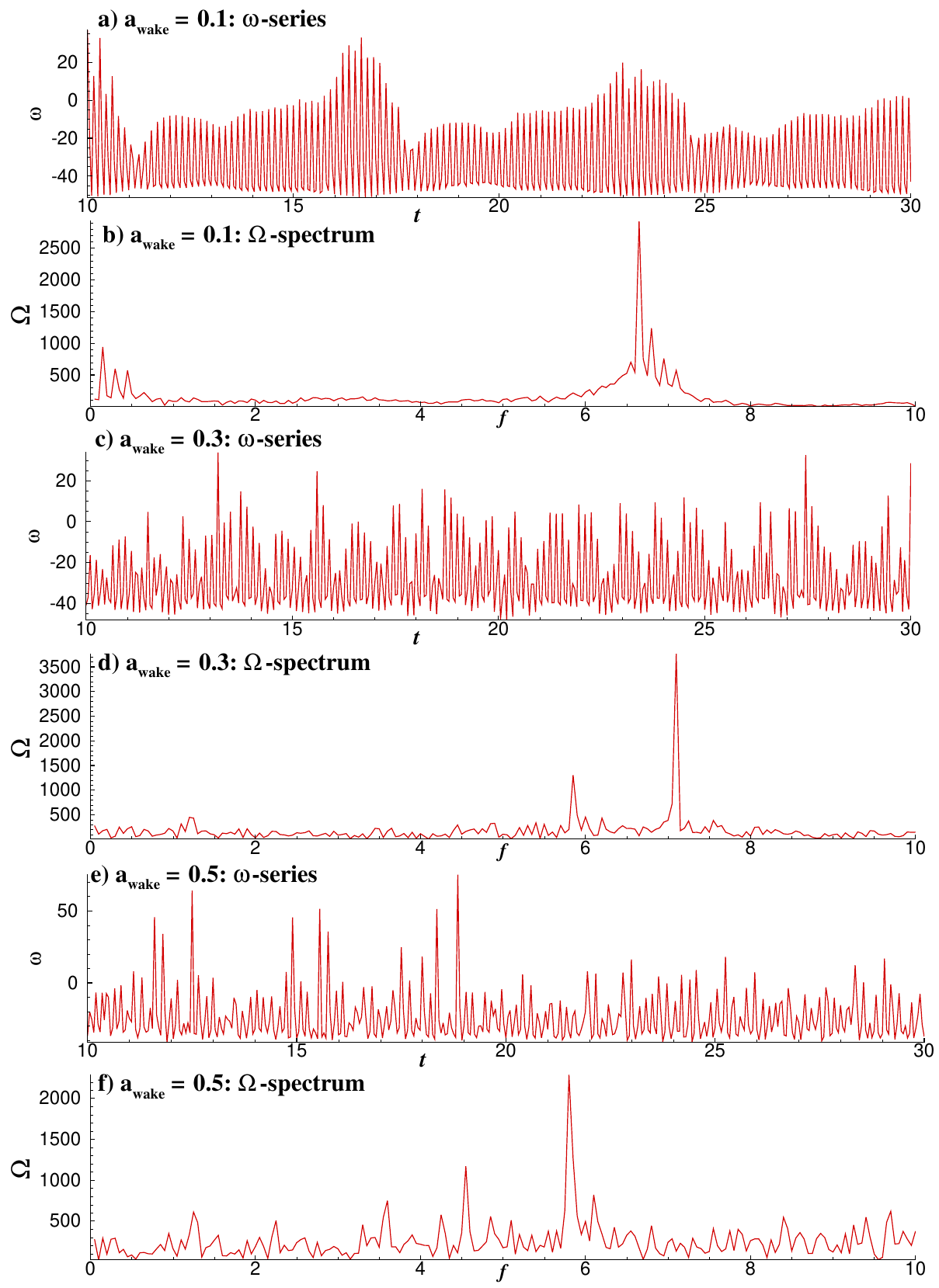}
\caption{Time-series of vorticity on the suction surface for inflow wake amplitudes (a) $a_{wake}$ = 0.1, (c) $a_{wake}$ = 0.3, (e) $a_{wake}$ = 0.5 and the corresponding spectra in frames (b), (d) and (f).}
\label{fig9}
\end{figure}

Increasing the wake amplitude alters vorticity dynamics, influencing vortical structures and separation bubbles, although this remains a quantitative comparison. To quantify these changes, we analyze the time-series of vorticity near the trailing edges of both suction and pressure surfaces and perform a Fast Fourier transform (FFT), providing the spectrum. In Fig. \ref{fig9}, the time series of the vorticity on the suction surface and corresponding FFT is shown for inflow wake amplitudes, $a_{wake} = 0.1$, 0.3 and 0.5. For low wake amplitude shown in Fig. \ref{fig9}(a), the time series reveals the presence of a dominant time period in the flow. This is confirmed in the spectrum of Fig. \ref{fig9}(b), where a peak is noted for $f \approx 7 Hz$, a sub-harmonic of the Kelvin-Helmholtz shedding frequency (14-16 Hz) \cite{mcauliffe2010transition}. The presence of a Kelvin-Helmholtz mechanism for uniform inflow has been established in the form of creation of 2D vortical rolls \cite{sengupta2020effectsa}. Upon increasing the wake amplitude to 0.3 in Fig. \ref{fig9}(c), the presence of multiple time periods becomes evident. The spectrum in Fig. \ref{fig9}(c) once again shows a dominant peak at $f = 7$, corresponding to the Kelvin-Helmholtz vortical rolls formed along the suction surface. In addition, a sub-dominant peak is noted at $f = 5.71 Hz$, which is a superharmonic of the wake passing frequency imposed at the inflow ($f_{wake} = 1/0.35 = 2.85 Hz$). Further increasing the wake amplitude to 0.5 in Fig. \ref{fig9}(e), displays a higher degree of chaos in the time-series and this is well-established in the spectrum of Fig. \ref{fig9}(f) where multiple peaks are noted across the frequency plane. The most dominant peak corresponds to the superharmonic of the wake passing frequency, i.e. at $f = 5.71Hz$, whereas the vorticity amplitude is distributed over a wide range of frequencies. This redistribution of vorticity over a larger range of frequencies is a feature that increases with wake amplitude, resulting in an overall reduction in the amplitude of $\Omega$. 

\begin{figure}[ht!]
\centering
\includegraphics[width=.9\textwidth]{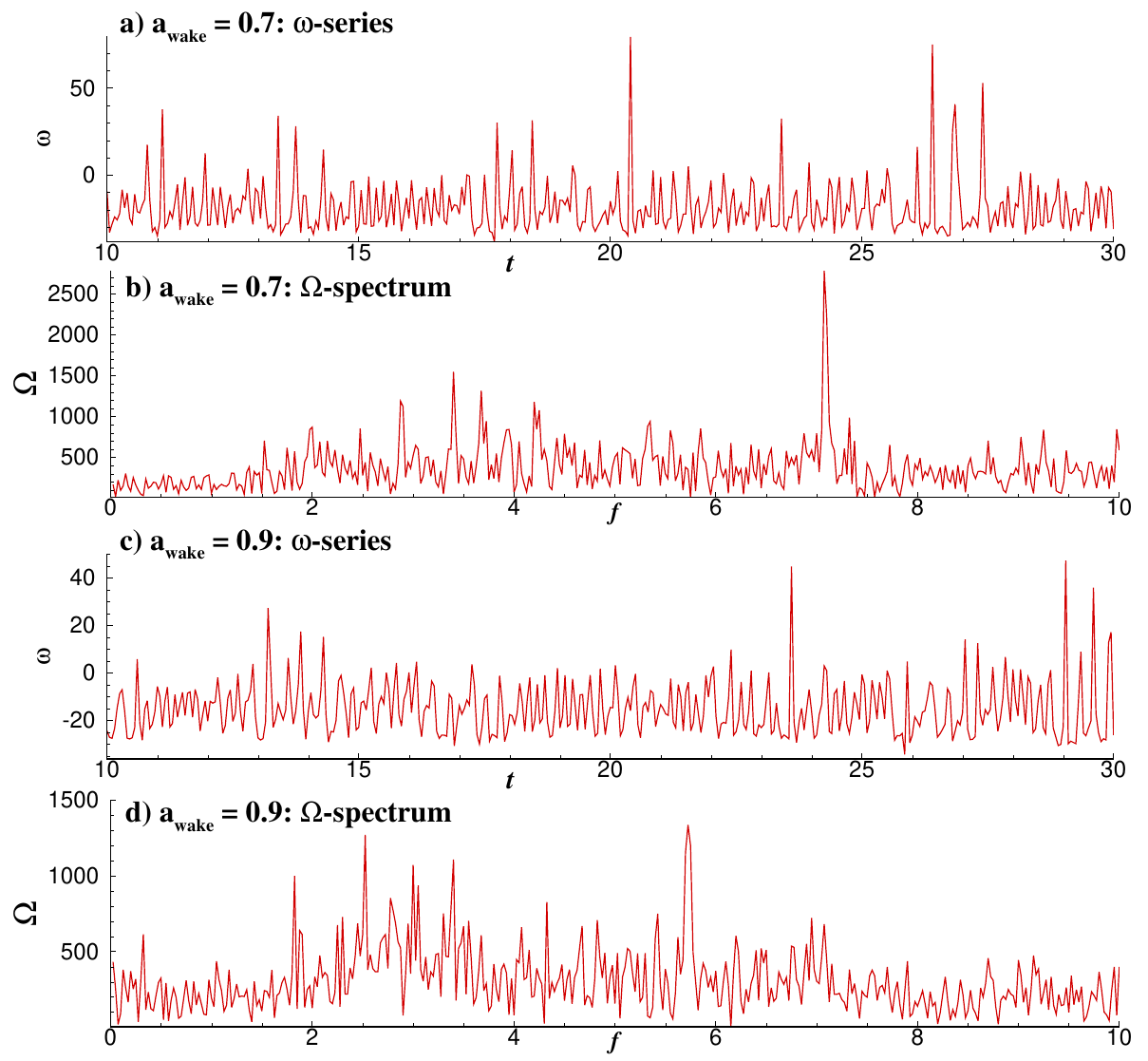}
\caption{Time-series of vorticity on the suction surface for inflow wake amplitudes (a) $a_{wake}$ = 0.7 and (c) $a_{wake}$ = 0.9 and the corresponding spectra in frames (b) and (d).}
\label{fig10}
\end{figure}

In Fig. \ref{fig10}, the time-series of vorticity and its spectrum on the suction surface are compared for two higher wake amplitudes, $a_{wake} = 0.7$ and 0.9. For the time-series corresponding to $a_{wake} = 0.7$, displayed in Fig. \ref{fig10}(a), the magnitude of vorticity further decreases, with multiple periods present in the time-series. Once again this reinforces the redistribution of vorticity over broader range of scales, which is indeed the case for the spectrum in Fig. \ref{fig10}(b). The maximum amplitude of $\Omega$ has also been found to decrease to 2500, while it decreases further for $a_{wake} = 0.9$ in Fig. \ref{fig10}(d) to 1500. As more frequencies gain prominence in the spectrum, there is no discernible peak present, rather multiple non-negligible peaks are observed. This complex, multi-periodic nature of the flow shows the interaction of two mechanisms: (i) one owing to the adverse pressure gradient forming on the suction surface and (ii) the other is due to the wake-induced transition which selectively amplifies certain low frequency components of the imposed wake due to a shear sheltering effect \cite{jacobs2001simulations}. The second of these is the reason for higher vorticity content at lower frequencies when $a_{wake}$ is increased beyond a value of 0.5. 

\begin{figure}[ht!]
\centering
\includegraphics[width=.9\textwidth]{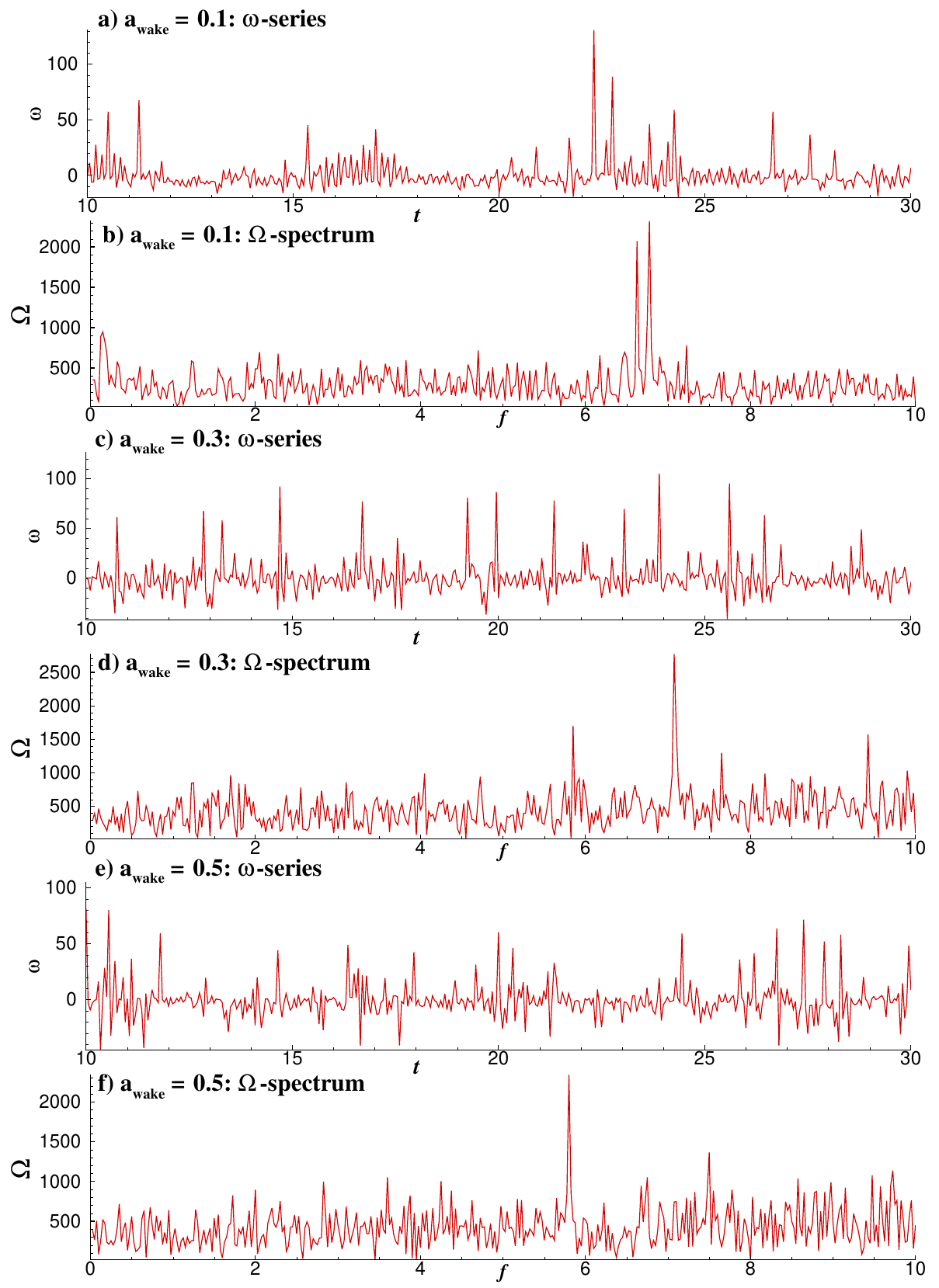}
\caption{Time-series of vorticity on the pressure surface for inflow wake amplitudes (a) $a_{wake}$ = 0.7 and (c) $a_{wake}$ = 0.9 and the corresponding spectra in frames (b) and (d).}
\label{fig11}
\end{figure}

In Fig. \ref{fig11}, the time series of the vorticity on the pressure surface and corresponding FFT is shown for inflow wake amplitudes, $a_{wake} = 0.1$, 0.3 and 0.5. As discussed in Fig. \ref{fig8}, the pressure surface reveals a stream of vortex shedding from the trailing edge. The time periods associated with individual vortical eddies are visible in the time series of Fig. \ref{fig11}(a) and in the corresponding spectrum of Fig. \ref{fig11}(b). The dominant peak in the spectrum is at a frequency that is a sub-harmonic of the Kelvin-Helmholtz shedding frequency \cite{mcauliffe2010transition}. In contrast to the suction surface, even for low wake amplitudes, the pressure surface displays a multi-periodic behavior. It is also to be noted that the overall magnitude of the vorticity is lower for the pressure surface than the suction surface. This can be attributed to the distribution of vorticity across various temporal scales during the vortex shedding on the pressure surface. On the suction surface, on the other hand, localized 2D vortical rolls are observed which introduce vorticity at fewer temporal scales, leading to a higher vorticity magnitude. For higher wake amplitude of 0.3 shown in Fig. \ref{fig11}(c), the magnitude of the vorticity reduces compared to $a_{wake} = 0.1$. This was also observed for Fig. \ref{fig8}, where increasing the wake amplitude led to the vortex shedding becoming less intense with lower vorticity magnitudes at the cores of the encompassing vortical eddies. The effect of the small-scale turbulence of the wakes in the free stream is to trigger the formation of turbulent spots \cite{emmons1951laminar}. In prior studies after a wake passing event, a period of \lq calm' with reduced turbulent activity \cite{schubauer1956contributions} was noted downstream of these turbulent spots, explaining the reduced vortex shedding intensity from the trailing edge of the pressure surface. The dominant peak for $a_{wake} = 0.3$ in Fig. \ref{fig11}(d) corresponds to Kelvin-Helmholtz frequency, which shifts to the superharmonic of the imposed wake passing frequency upon increasing the amplitude to 0.5 in Fig. \ref{fig11}(f), as noted for the suction surface in Fig. \ref{fig9}.

\begin{figure}[ht!]
\centering
\includegraphics[width=.9\textwidth]{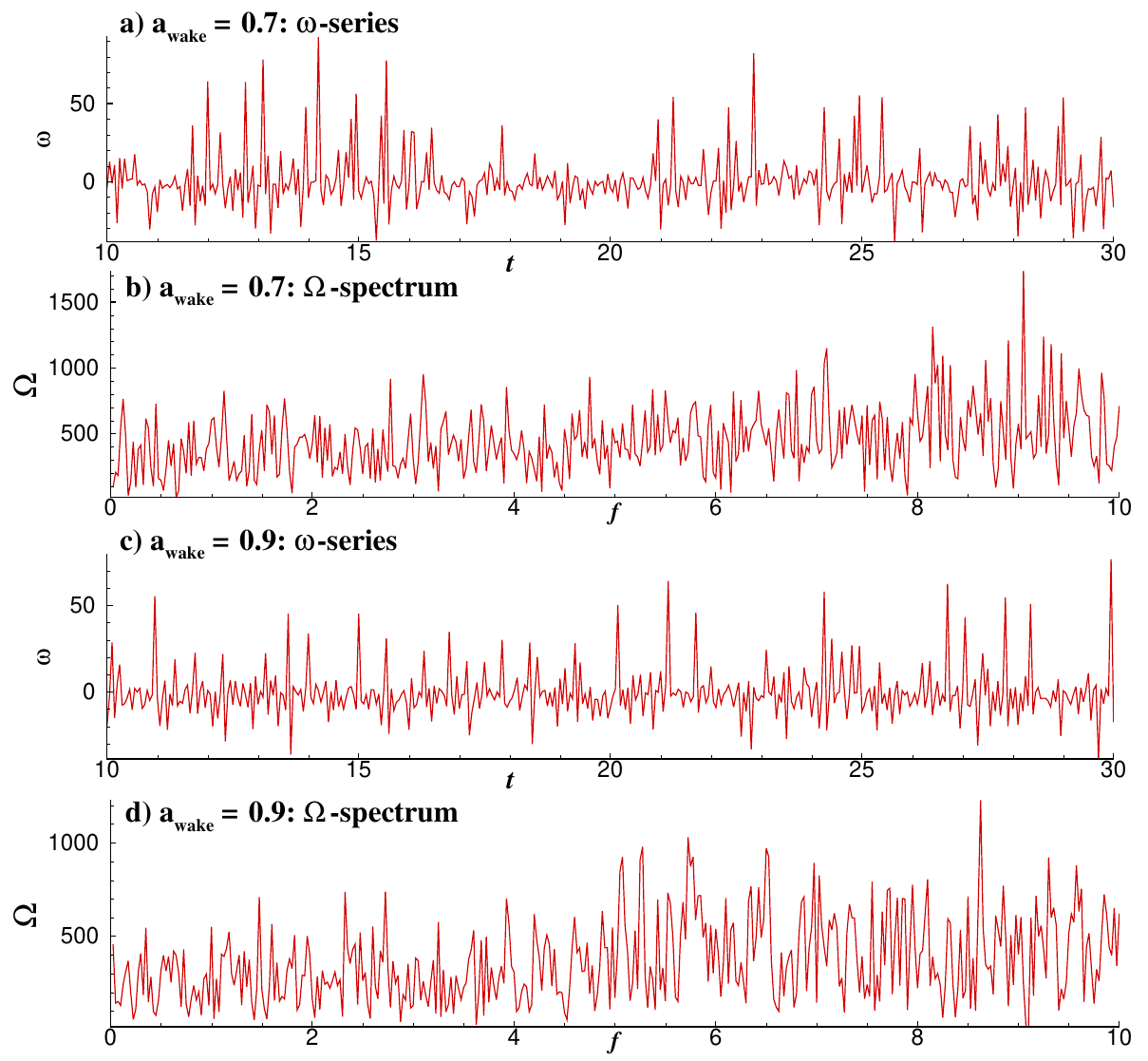}
\caption{Time-series of vorticity on the pressure surface for inflow wake amplitudes (a) $a_{wake}$ = 0.7 and (c) $a_{wake}$ = 0.9 and the corresponding spectra in frames (b) and (d).}
\label{fig12}
\end{figure}

In Fig. \ref{fig12}, the time-series of vorticity and its spectrum on the pressure surface are compared for two higher wake amplitudes, $a_{wake} = 0.7$ and 0.9. The spectra for both these wake amplitudes do not show any discernible peak in the frequency plane. Rather, the amplitude of the vorticity is reduced significantly, with redistribution of vorticity across a broad range of temporal scales. This chaotic, multi-periodic nature of the spectrum is characteristic of soft turbulence, observed for another internal flow triggered by a separated shear layer \cite{joshi2023exploring}. 

\begin{figure}[ht!]
\centering
\includegraphics[width=.88\textwidth]{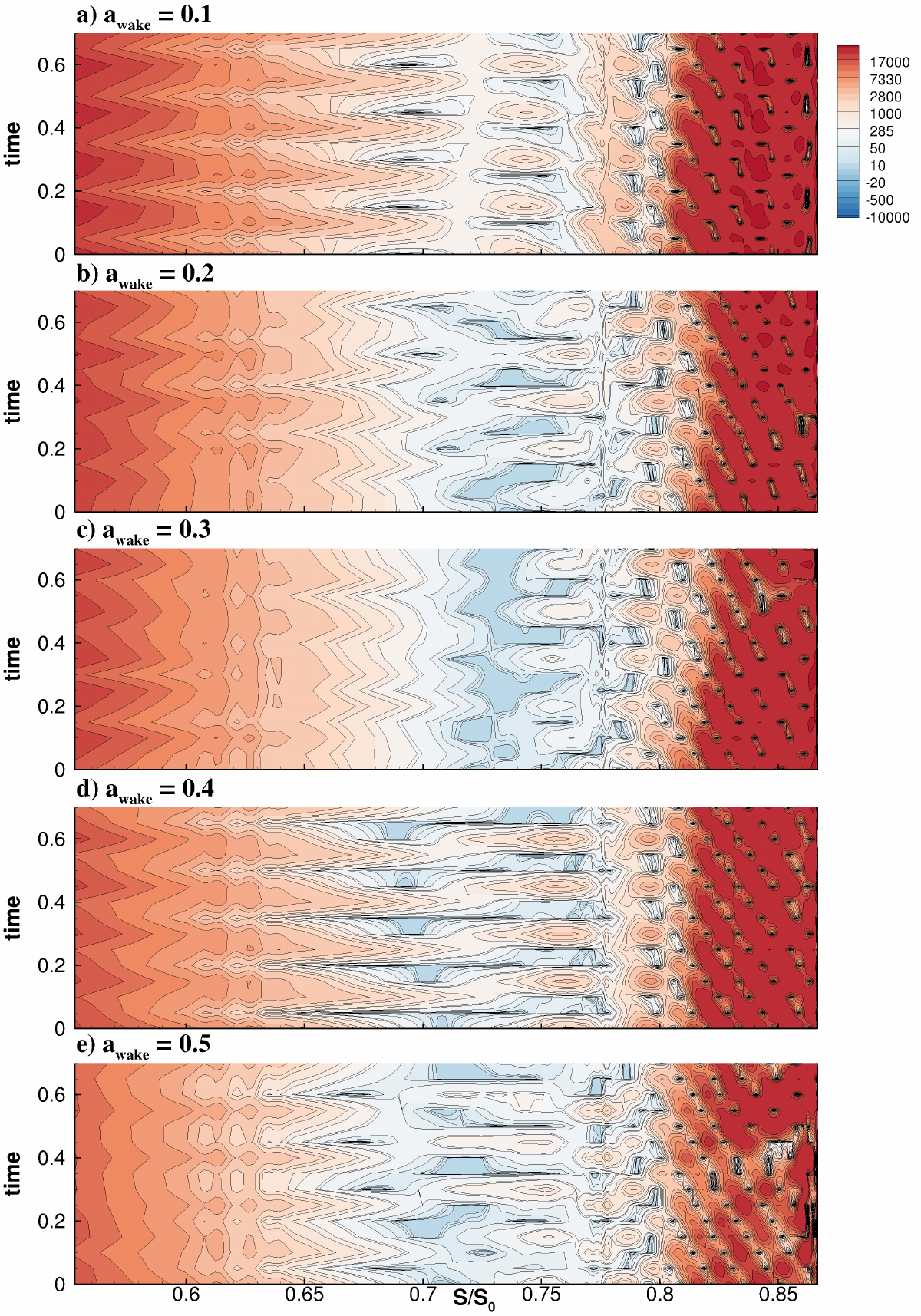}
\caption{Space-time plots of enstrophy for inflow wake amplitudes, (a) $a_{wake}$ = 0.1, (b) $a_{wake}$ = 0.2, (c) $a_{wake}$ = 0.3, (d) $a_{wake}$ = 0.4 and (e) $a_{wake}$ = 0.5.}
\label{fig13}
\end{figure}

\begin{figure}[ht!]
\centering
\includegraphics[width=.88\textwidth]{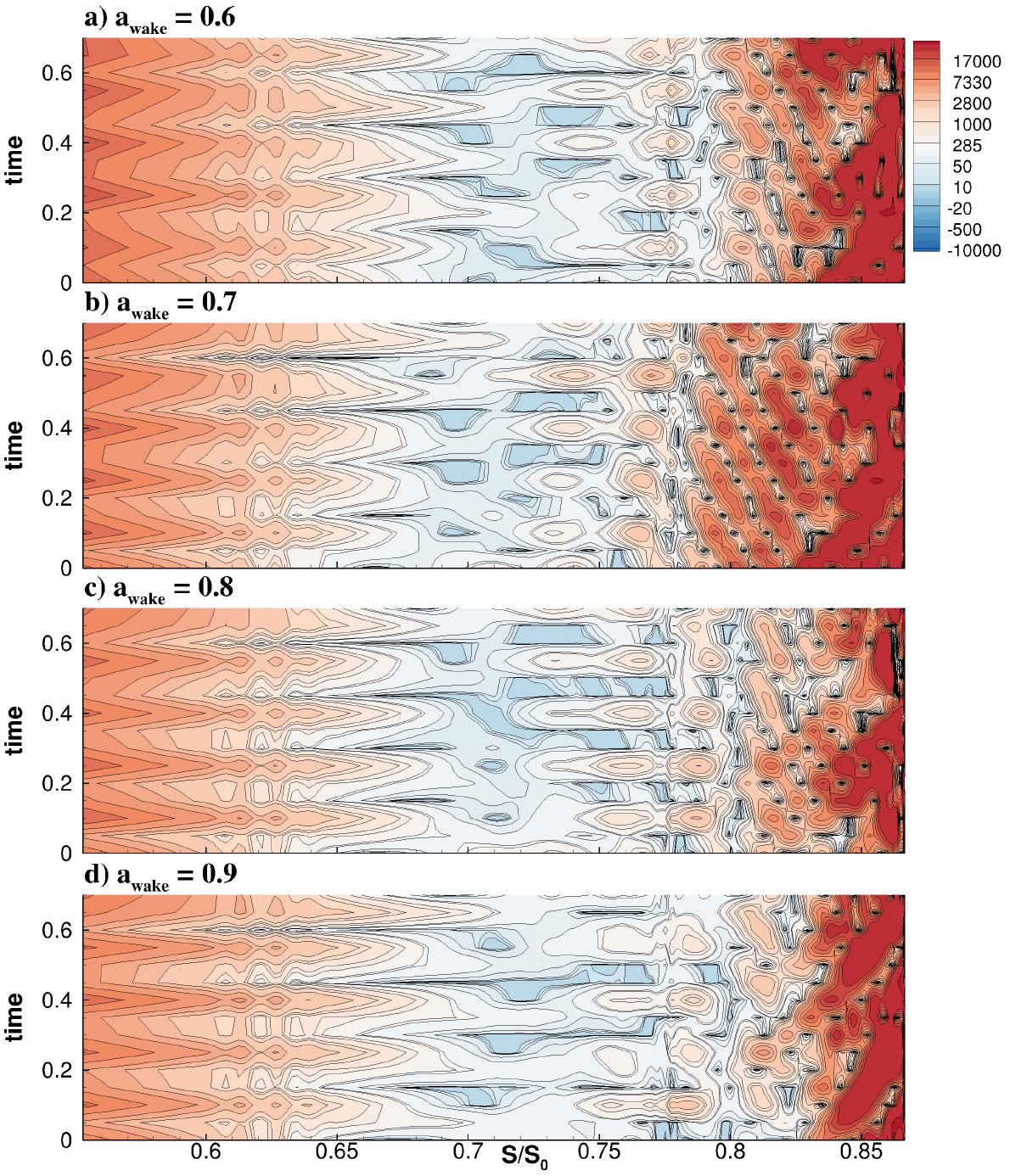}
\caption{Space-time plots of enstrophy for inflow wake amplitudes, (a) $a_{wake}$ = 0.6, (b) $a_{wake}$ = 0.7, (c) $a_{wake}$ = 0.8 and (d) $a_{wake}$ = 0.9.}
\label{fig14}
\end{figure}

Next, we calculate the enstrophy, $\Omega_c$ as a measure of the rotational energy produced in the flow field subjected to incoming wakes of varying amplitudes, and project it in the form of a space-time diagram in Figs. \ref{fig13} and \ref{fig14}. These are useful in observing the variation of the boundary layer parameters over the wake-passing cycle. The height at which this plot is produced is $y/S_0 = 0.0252$, which is in the immediate vicinity of the suction surface. 

In Fig. \ref{fig13}, the space-time plots of enstrophy are compared for incoming wakes with amplitudes, $a_{wake} = 0.1$ to 0.5, over two time periods of the imposed wake ($t_{wake} = 0.35$). These can be considered cases of wake-induced transition with a relatively low amplitude of imposed wakes at the inflow plane. For low wake amplitudes, shown in Figs. \ref{fig13}(a) to (c), for $S/S_0 < 0.6$, longitudinal \lq puffs' are observed in the boundary layer during the receptivity phase \cite{wu1999simulation} of the flow. These puffs stretch into streaky structures (noted at $S/S_0 > 0.65$) as they convect downstream which are more prominent for the higher wake amplitudes displayed in Figs. \ref{fig13}(d) and \ref{fig13}(e). These puffs further breakdown into turbulent spots \cite{emmons1951laminar}. Turbulent spots spread at a constant angle and have a wedge-like appearance. For $a_{wake} > 0.2$, downstream of the streaks and turbulent spots one notes regions of low enstrophy magnitude, indicating a lower rotationality. This corresponds to the \lq calmed region' having reduced turbulence \cite{coull2012predicting}, and hence reduced generation of vortical eddies. For the highest wake amplitude shown here in Fig. \ref{fig13}(e), we measure the speed at which the turbulent spot convects near its leading edge. This is found to be 83.34\% of the free stream velocity, matching with Schubauer and Klebanoff's measurements \cite{schubauer1956contributions} of the leading edge turbulent spot velocity of 88\%. 

In Fig. \ref{fig14}, the space-time plots of enstrophy are compared for incoming wakes with amplitudes, $a_{wake} = 0.6$ to 0.9, over two time periods of the imposed wake. For these higher wake amplitudes, the \lq calmed region' downstream of the turbulent spot is spread over a larger streamwise extent, marked by lower $\Omega_c$. The maximum region with reduced enstrophy is observed for the highest wake amplitude, $a_{wake} = 0.9$. In the classical experiments of Gostelow {\it et al.} \cite{gostelow1997investigation} the convection speed of the calmed region was measured as 20-30\% of free-stream velocity. Here, the speed of structures noted in the calmed region is calculated from Fig. \ref{fig14}(d) and found to be 25.86 \% of the free stream velocity. The turbulent spot is found to elongate as it convects downstream, due to varying speeds of the leading edge and trailing edge of the spot. The trailing edge convects at 50\% of the free stream speed while the leading edge is convecting at 83.3\% of the free stream speed \cite{schubauer1956contributions}. Thus, incoming wakes triggered an earlier wake-induced transition. The resultant turbulent spots and associated calmed regions convected downstream and effectively suppressed the separation bubble. The periodic suppression of the
separation bubble due to the wake has a benefit in reducing the blade's profile loss, especially for low $Re$.

\subsection{Energy budget for wake-induced transition in a T106A LPT passage}

The turbulent kinetic energy (TKE) budget is crucial in fluid dynamics, particularly in understanding and modeling turbulence. The TKE budget describes the production, transport, and dissipation of kinetic energy in turbulent flows. It is a measure of the translational energy, which is typically produced by the work done by mean flow gradients on the turbulent fluctuations. This is critical in characterizing turbulent regions, particularly in boundary layers. Here, the TKE budget provides insights into the mechanisms of energy production near walls, which is crucial for predicting drag. The budget for TKE is written as:

\begin{equation}
\frac{\partial k}{\partial t} + C = P - \epsilon - (TD + PD + VD)
\label{tke}
\end{equation}

\noindent where, $P = -u^{'}_{i} u^{'}_{j} \frac{\partial U_i}{\partial x_j}$ is the TKE production term, $C$ is convection and $\epsilon$ is the dissipation. The terms $u^{'}$ are the fluctuating components of the velocity, while $U$ is the time-averaged mean velocity component. The net diffusion is provided by the interplay between turbulent diffusion (TD), pressure diffusion (PD), and viscous diffusion (VD). Here, we will limit our focus to TKE production alone, the complete relations have been provided by Alam and Sandham \cite{alam2000direct}. 

\begin{figure}[ht!]
\centering
\includegraphics[width=.88\textwidth]{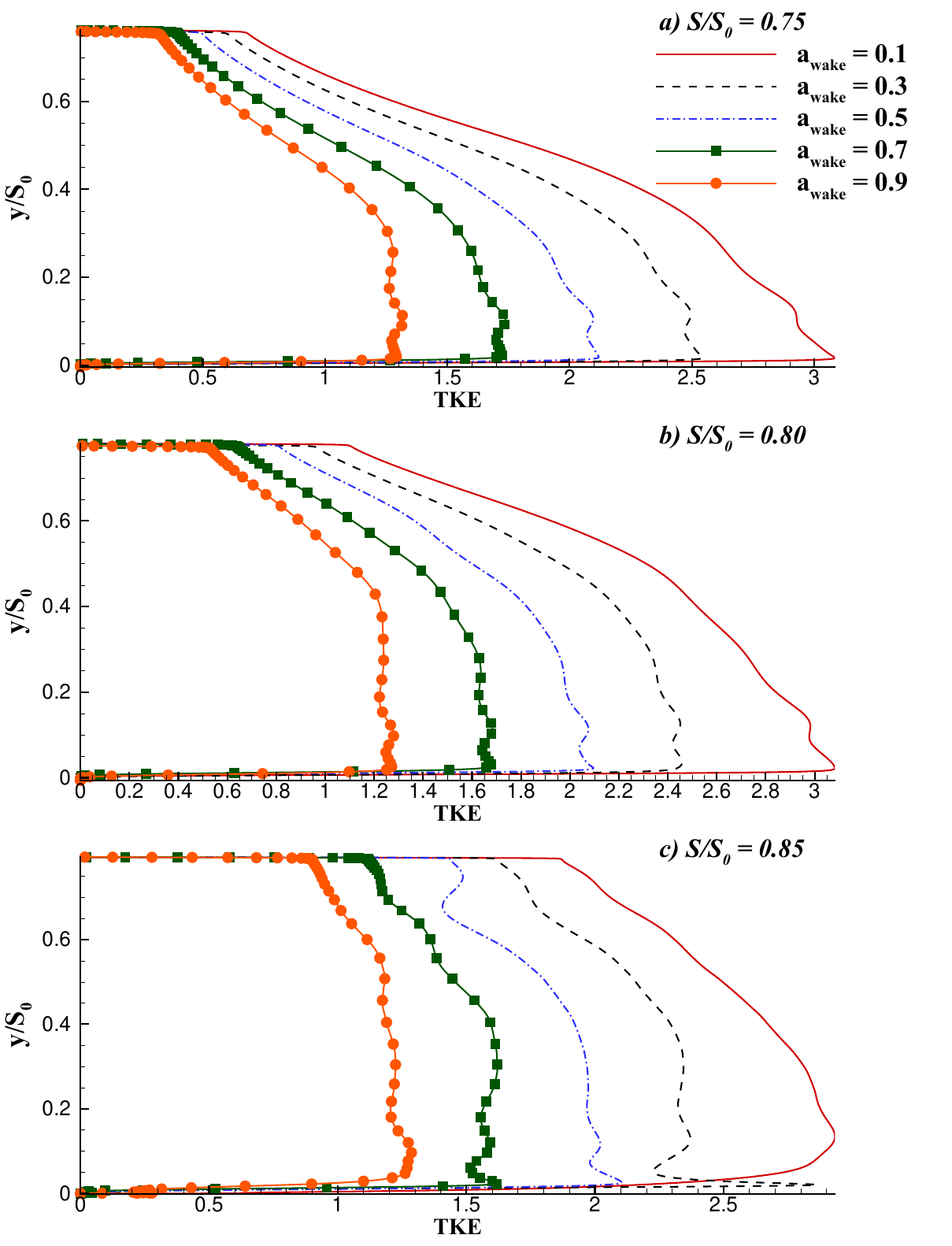}
\caption{Wall-normal distribution of TKE production for inflow wake amplitudes, $a_{wake}$ = 0.1, 0.3, 0.5, 0.7 and 0.9, extracted at (a) $S/S_0 = 0.75$, (b) $S/S_0 = 0.80$ and (c) $S/S_0 = 0.85$.}
\label{fig15}
\end{figure}

In Fig. \ref{fig15}, the wall-normal distribution of TKE production is compared for the indicated inflow wake amplitudes for three critical locations along the suction surface: (a) $S/S_0 = 0.75$  which is upstream of flow separation location, (b) $S/S_0 = 0.80$ which is downstream of flow separation, and (c) $S/S_0 = 0.85$ which coincides with flow reattachment location. For all three locations, with increasing $a_{wake}$, the production of TKE decreases. This can be explained by the \lq calmed regions' formed by the wake-induced transition which periodically suppress the turbulent activity - an aspect that is enhanced by increasing the wake amplitude. Enhanced mixing is noted with the wall-normal location of the peak TKE shifting closer to the wall \cite{alam2000direct} with an increase in $a_{wake}$. New peaks in TKE production are observed in the free stream due to the small-scale turbulence of the wakes \cite{wu1999simulation}, which is more prominent for higher wake amplitudes. Flow at these locations possess characteristics of free shear layers and a newly formed turbulent boundary layer \cite{sengupta2020effects}. Comparing the locations along the suction surface, one concludes that the peak TKE production is immediately after flow separation in Fig. \ref{fig15}(b), followed by the pre-separated flow in Fig. \ref{fig15}(a). The least TKE production is for the location post-reattachment in Fig. \ref{fig15}(c), due to turbulent dissipation becoming prominent after the shear layer starts to reattach \cite{sengupta2017roughness}. 

The wall-normal locations of the peak TKE production shown in Fig. \ref{fig15} for three locations along the suction surface are tabulated for various wake amplitudes in Table \ref{tab3}. This establishes that the peak shifts towards the wall with an increase in wake amplitude, due to enhanced mixing for all primary, secondary, and tertiary TKE productions. Post reattachment, the flow experiences tertiary TKE production due to small-scale turbulence introduced by the Gaussian wakes in the free stream, recorded as $y_{p3}$. As one progresses from the pre-separated to post-reattachment flows, the primary TKE peaks shift away from the wall. This suggests that the turbulence is produced due to the wakes in the free stream for these downstream locations, and not from within the boundary layer.

\begin{table}[h!]
\centering
\caption{Wall-normal locations of peak (primary, secondary, and tertiary) TKE production at $S/S_0 = 0.75$, 0.80 and 0.85, for different inflow wake amplitudes.}
\vspace{1mm}
\begin{tabular}{|c| c| c| c| c| c| c| c| c| c|}
\hline
 $a_{wake}$ & $S/S_0=0.75$:$y_{p1}$ & $y_{p2}$  & $S/S_0=0.80$:$y_{p1}$  & $y_{p2}$  & $y_{p3}$  & $S/S_0=0.85$:$y_{p1}$ & $y_{p2}$ & $y_{p3}$ \\ [0.5ex]
 \hline\hline
    0.1 & 0.0159 & - & 0.0476 & - & - & 0.1358 & - & - \\ 
    \hline
    0.2 & 0.0155 & 0.1217 & 0.0372 & - & - & 0.1254 & - & - \\ 
    \hline
    0.3 & 0.0152 & 0.1181 & 0.0313 & 0.1249 & - & 0.1207 & - & - \\
    \hline
    0.4 & 0.0148 & 0.1133 & 0.0270 & 0.1242 & - & 0.1189 & 0.0312 & - \\
    \hline
    0.5 & 0.0145 & 0.1061 & 0.0251 & 0.1187 & - & 0.1163 & 0.0280 & - \\
    \hline
    0.6 & 0.0138 & 0.0971 & 0.0234 & 0.1143 & 0.3043 & 0.1130 & 0.0254 & - \\
    \hline
    0.7 & 0.0132 & 0.0953 & 0.0215 & 0.1136 & 0.2737 & 0.1112 & 0.0238 & 0.3086 \\
    \hline
    0.8 & 0.0127 & 0.0934 & 0.0203 & 0.1056 & 0.2605 & 0.1057 & 0.0206 & 0.2681 \\
    \hline
    0.9 & 0.0119 & 0.0913 & 0.0195 & 0.1031 & 0.2484 & 0.1003 & 0.0191 & 0.2408 \\
    [1ex]
 \hline
\end{tabular}

\label{tab3}
\end{table}

Like TKE is a measure of translational energy, we have compressible enstrophy ($\Omega_c$) as an effective measure of rotational energy in the flow. A compressible enstrophy transport equation (CETE) has been developed \cite{suman2022novel} to explain the underlying mechanisms in the creation, growth, and redistribution of enstrophy. Since the flow during the wake-induced transition is dominated by coherent vortical structures on the suction and pressure surfaces, it is imperative to analyze the effect that the wake amplitude has on the enstrophy growth and thereby on the instability. As enstrophy is always a positive quantity, a positive growth rate of enstrophy is an indicator of flow instability. Previous applications of the CETE to flows dominated by pressure gradients \cite{sengupta2019direct, sengupta2023compressibility} revealed that for a 2D flow, the prominent mechanism features an interplay between baroclinic and viscous stress terms. For free shear layers \cite{sengupta2023multi} dominated by density gradients, the baroclinic term and vortex stretching mechanisms become important during various stages of flow instability in 3D. The vortex stretching term is notably absent for 2D flows. The CETE has not been applied to the flow inside a T106A LPT passage affected by wake-induced transition, and here we will show the mechanisms for enstrophy transport due to increasing wake amplitude. The budget terms of the CETE \cite{suman2022novel} are as follows:

\begin{equation}
\begin{aligned}
 \frac{D{\Omega_c }}{Dt} = & 2\vec{\omega } \cdot \left[(\vec{\omega} \cdot {\nabla}) \vec{V}\right] - 2({\nabla} \cdot \vec{V}) \Omega_c + \left(\frac{2}{\rho^{2}}\right) \vec{\omega } \cdot \left[\left({\nabla \rho} \times {{\nabla p}}\right)\right] +\left(\frac{4}{\rho}\right)\vec{\omega } \cdot \left[{\nabla} \times \left[{\nabla} \cdot \left(\mu S \right)\right]\right]  \\\\
 & -\left(\frac{2}{\rho^{2}}\right)\vec{\omega } \cdot \left[{\nabla \rho} \times {\nabla} \left(\lambda ({\nabla} \cdot \vec{V})\right)\right] -\left(\frac{4}{\rho^2}\right)\vec{\omega } \cdot \left({\nabla \rho} \times ({\nabla} \cdot \left(\mu S\right)) \right) 
\end{aligned}
\label {MderCETE_final}
\end{equation}

Here, $\vec{\omega }$ and $\vec{V}$ represent the vorticity and velocity vectors, and $S$ is the strain rate tensor. The various terms of the CETE budget are described as follows: \\[1.0ex]

$2\vec{\omega } \cdot \left[(\vec{\omega} \cdot {\nabla}) \vec{V}\right]$ : Enstrophy contribution from vortex stretching (T1).\\[1.1ex]

$({\nabla} \cdot \vec{V}) \Omega_c$: Enstrophy growth/decay due to compressibility (T2). \\ [1.1ex]

$\left(\frac{1}{\rho^{2}}\right) \vec{\omega } \cdot \left[\left({\nabla \rho} \times {{\nabla p}}\right)\right]$: Baroclinic contribution due to misalignment of gradients of pressure and density (T3).\\[1.1ex]

$\left(\frac{1}{\rho^{2}}\right)\vec{\omega } \cdot \left[{\nabla \rho} \times {\nabla} \left(\lambda ({\nabla} \cdot \vec{V})\right)\right]$: Contribution due to misalignment of vorticity and bulk viscosity gradients (T4) .\\[1.1ex]

$\left(\frac{1}{\rho}\right)\vec{\omega } \cdot \left[{\nabla} \times \left[{\nabla} \cdot \left(\mu S \right)\right]\right]$: Diffusion of enstrophy due to viscous action (T5).\\[1.1ex]

$\left(\frac{1}{\rho^2}\right)\vec{\omega } \cdot \left({\nabla \rho} \times ({\nabla} \cdot \left(\mu S\right)) \right)$: Contribution due to misalignment of gradients of density and divergence of viscous stresses (T6).\\[1.1ex]

\begin{figure}[ht!]
\centering
\includegraphics[width=.9\textwidth]{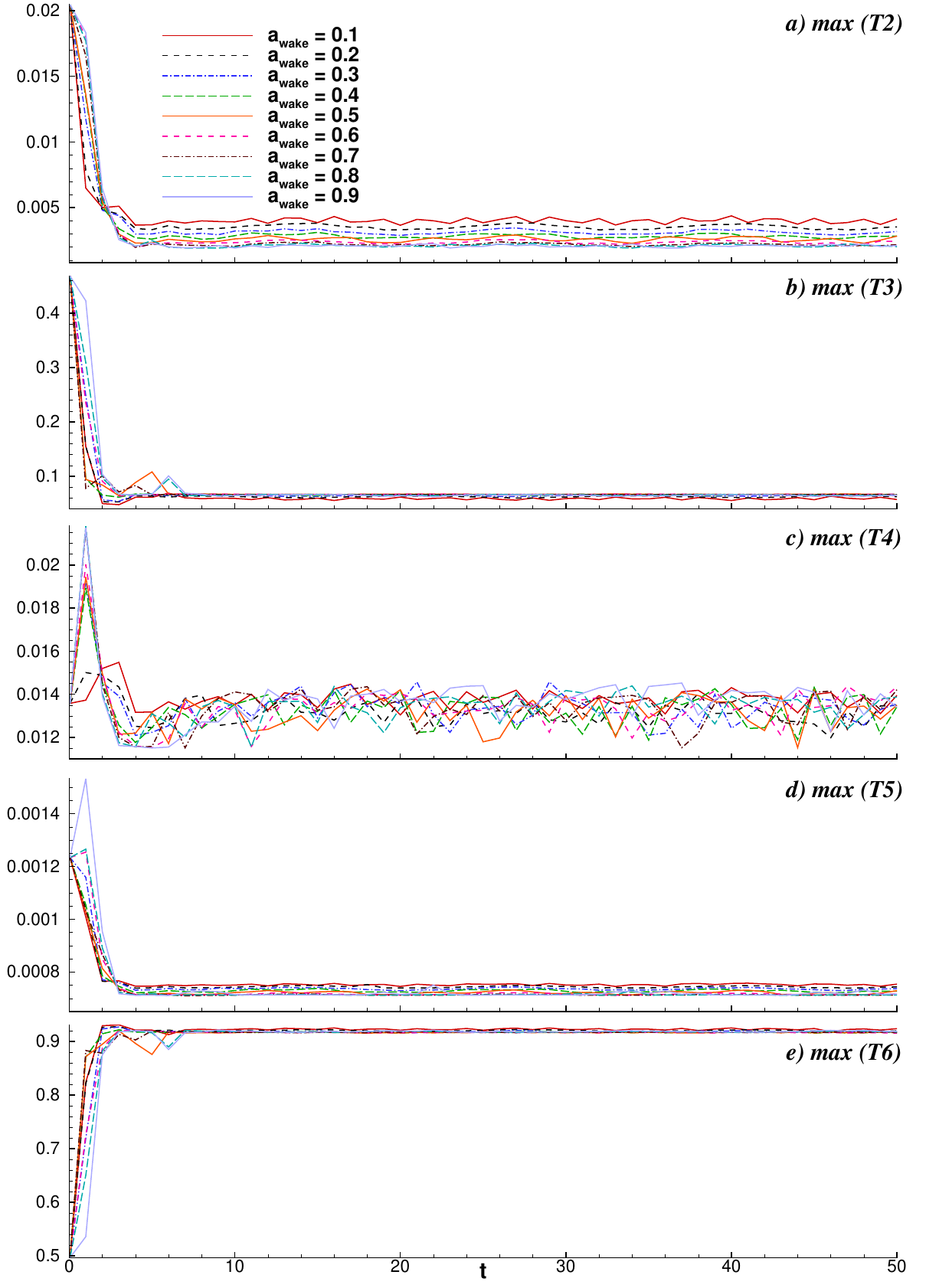}
\caption{Evolution of maximum CETE budget terms (a) T2, (b) T3, (c) T4, (d) T5 and (e) T6; for inflow wake amplitudes, $a_{wake}$ = 0.1 to 0.9.}
\label{fig16}
\end{figure}

In Fig. \ref{fig16}, the evolution of the maximum fractional contribution of CETE budget terms of Eq. \eqref{MderCETE_final} are shown for inflow wake amplitudes ranging from 0.1 to 0.9. Irrespective of the $a_{wake}$, the largest contribution to CETE is from T6, which is owing due to the viscous stress term. The second largest contribution is from T3, which is the baroclinic term. These observations are consistent with prior applications of the CETE to the T106A passage studying role of Mach number \cite{sengupta2023compressibility} and free stream turbulence \cite{sengupta2024separation}. As the flow field is dominated by the formation of vortical rolls and vortex shedding from suction and pressure surfaces, respectively, the contribution from baroclinic vorticity is high. When $a_{wake}$ is increased, the contributions from terms T3, T4, and T5 during the onset increase, resulting in an overall reduction of contribution from T6. This can be explained by the addition of the negative jet \cite{addison1990unsteady} via the imposed wakes, which directly influence the velocity and vorticity fields. Since the operating Mach number is 0.15, the effect of T2 (due to compressibility) is not significant, with marginal changes noted with increasing $a_{wake}$. Apart from term T4 (due to misalignment of vorticity and bulk viscosity gradient), the wake amplitude does not play a significant role during the later stages of the flow instability. The trend of the CETE with wake amplitude is quite straightforward. However, the frequency of wake passing is likely going to be more receptive to competing mechanisms in the CETE budget, and should serve as the focus for future studies involving CETE with wake-induced transition. 

\section{Summary and Conclusions}\label{sec4}

The current study focuses on the effects of varying wake amplitudes on the flow characteristics within a T106A low-pressure turbine (LPT) passage. Wake-induced transition plays a significant role in altering boundary layer parameters, vorticity dynamics, turbulent kinetic energy (TKE) production, and compressible enstrophy ($\Omega_c$) within the flow. By solving the two-dimensional compressible Navier-Stokes equations over nine amplitudes of the incoming periodic Gaussian wakes, several key findings are highlighted. 

Increasing wake amplitude leads to changes in boundary layer profiles, influencing maximum velocity, momentum thickness, and separation bubble characteristics along the suction surface. Higher wake amplitudes mitigate adverse pressure gradients, reducing peak velocities and suppressing the net region subjected to flow separation. For instance, a ninefold increase in wake amplitude leads to a 50\% reduction in the skin friction drag, a 39.2\% increase in momentum thickness, and a 37\% increase in bubble height. This in turn, indicates that increased wake amplitude has a potential benefit for reducing the profile loss of a T106A blade.

The characteristic features of wake-induced transition feature the formation of coherent structures (such as puffs, streaks, turbulent spots and calmed regions) inside the boundary layer and in the free stream, as depicted in the space-time plots. The small-scale turbulence of the imposed wakes in the free stream in particular, transfers turbulent fluid to the edge of the boundary layer, which affects the vorticity dynamics. Increasing wake amplitudes generally decrease vorticity magnitudes near the suction surface, redistributing vorticity over multiple separation bubbles. On the pressure surface also, the intensity of the vortex shedding reduces with wake amplitude due to the small-scale turbulence of the wakes in the free stream. 

The production of turbulent kinetic energy (TKE) decreases with increasing wake amplitude along the suction surface. As wake-induced transition invokes intermittent calmed regions (with reduced turbulent activity) downstream of turbulent spot formation, it leads to a reduction of the turbulence in the flow. Peak TKE production shifts closer to the wall, indicating enhanced mixing with increase in wake amplitude. New TKE productions are noted in free stream for higher wake amplitudes, influenced by small-scale turbulence (imposed by Gaussian wakes). The application of a compressible enstrophy transport equation (CETE) shows dominant contributions from viscous stress terms, followed by baroclinicity. The contribution from the latter increases with an increase in wake amplitude, possibly due to the creation of more number of unsteady separation bubbles along the suction surface. 

Future research should focus on the frequency of wake passing as it will be more receptive to the competing mechanisms governing the wake-induced transition. Further numerical investigations could provide deeper insights into optimizing turbine blade designs for enhanced performance under varying flow conditions, particularly by extending the study to 3D. Overall, the study underscores the critical role of wake-induced transition in altering flow characteristics within turbine passages. It provides a foundation for optimizing turbine blade designs to minimize losses and improve efficiency in practical applications keeping the wake amplitude as the optimizing criterion. 

 \section*{Credit authorship contribution statement}
Aditi Sengupta: Conceptualization and Supervision, Methodology, Formal analysis, Funding acquisition, Investigation, Project administration, Resources, Writing- Original draft preparation. 

\section*{Declaration of competing interest}
The author declares that they have no known competing financial interests or personal relationships that could have appeared to influence the work reported in this paper.

\section*{Data Availability}
The data that support the findings of this study are available from Aditi Sengupta (aditi@iitism.ac.in) upon reasonable request.

\section*{Acknowledgements}
The authors acknowledge the use of the high-performance computing facility, ARYABHATA at IIT (ISM) Dhanbad for computing all the cases reported here. The authors are grateful to the Department of Science and Technology (DST), Government of India, Grant No. SRG/2022/0079 and IIT (ISM) Grant No. FRS(162)/2021-2022/MECH for providing full financial support for the problem solved here.
	
\bibliographystyle{elsarticle-num} 
\bibliography{wake_transition}
	
\end{document}